\documentclass[fleqn,usenatbib]{mnras}
\usepackage{newtxtext,newtxmath}

\usepackage[T1]{fontenc}

\DeclareRobustCommand{\VAN}[3]{#2}
\let\VANthebibliography\thebibliography
\def\thebibliography{\DeclareRobustCommand{\VAN}[3]{##3}\VANthebibliography}

\usepackage{graphicx}	
\usepackage{amsmath}	
\usepackage{threeparttable}
\usepackage{xcolor}

\title[MOJAVE--\textit{Fermi}: $\gamma$-ray emission region in AGN]{A decade of joint MOJAVE--\textit{Fermi} AGN monitoring: localisation of the gamma-ray emission region}

\author[Kramarenko et al.]{\parbox{\textwidth}{
I. G. Kramarenko,$^{1}$\thanks{E-mail: im.kramarenko@gmail.com}
A. B. Pushkarev,$^{3,2,1}$\thanks{E-mail: pushkarev.alexander@gmail.com}
Y. Y. Kovalev,$^{2,1,4}$
M. L. Lister,$^{5}$
T. Hovatta,$^{6,7}$
T. Savolainen$^{8,7,4}$
}
\vspace{0.4cm}\\
\parbox{\textwidth}{
$^{1}$Moscow Institute of Physics and Technology, Institutsky per.~9, Dolgoprudny, Moscow region, 141700, Russia\\
$^{2}$Lebedev Physical Institute of the Russian Academy of Sciences,
Leninsky prospekt 53, 119991 Moscow, Russia\\
$^{3}$Crimean Astrophysical Observatory, Nauchny 298688, Crimea, Russia\\
$^{4}$Max-Planck-Institut f\"ur Radioastronomie, Auf dem H\"ugel 69, 53121 Bonn, Germany\\
$^{5}$Department of Physics and Astronomy, Purdue University, 525 Northwestern Avenue, West Lafayette, IN 47907, USA\\
$^{6}$Finnish Centre for Astronomy with ESO, University of Turku, FI-20014 Turku, Finland\\
$^{7}$Aalto University Mets\"ahovi Radio Observatory, Mets\"ahovintie 114, FI-02540 Kylm\"al\"a, Finland\\
$^{8}$Aalto University Department of Electronics and Nanoengineering, PL 15500, FI-00076 Aalto, Finland
}}

\date{Accepted 2021 November 15. Received 2021 November 9; in original form 2021 June 15}

\pubyear{2021}

\begin{document}
\label{firstpage}
\pagerange{\pageref{firstpage}--\pageref{lastpage}}
\maketitle

\begin{abstract}
Within the MOJAVE VLBA program (Monitoring of Jets in AGN with VLBA Experiments), we have accumulated observational data at 15~GHz for hundreds of jets in $\gamma$-ray bright active galactic nuclei since the beginning of the \textit{Fermi} scientific observations in August 2008. We investigated a time delay between the flux density of AGN parsec-scale radio emission at 15 GHz and 0.1--300~GeV \textit{Fermi} LAT photon flux, taken from constructed light curves using weekly and adaptive binning. The correlation analysis shows that radio is lagging $\gamma$-ray radiation by up to 8~months in the observer's frame, while in the source frame, the typical delay is about 2-3~months. If the jet radio emission, excluding the opaque core, is considered, significant correlation is found at greater time lags.
We supplement these results with VLBI kinematics and core shift data to conclude that the dominant high-energy production zone is typically located at a distance of several parsecs from the central nucleus.
We also found that quasars have on average more significant correlation peak, more distant $\gamma$-ray emission region from the central engine and shorter variability time scale compared to those of BL~Lacertae objects.
\end{abstract}

\begin{keywords}
galaxies: active -- galaxies: nuclei -- galaxies: jets -- gamma-rays: galaxies -- radio continuum: galaxies
\end{keywords}



\section{Introduction}

The question of the dominant production mechanism and the exact location of the $\gamma$-ray emission observed in active galactic nuclei (AGNs) remains unresolved for several decades. The lack of high-resolution instruments and the complex nature of AGNs have resulted in numerous hypotheses about the origin of the $\gamma$-ray photons.

Since the era of \textit{CGRO}/EGRET \citetext{{\it Compton Gamma-Ray Observatory}/Energetic Gamma-Ray Experiment Telescope, \citealp{1999ApJS..123...79H}}, it has become clear that the high-energy emission is strongly associated with jet activity \citep[e.g.][]{1992Sci...257.1642D, 1995ApJ...440..525V}, which was later successfully confirmed by \textit{Fermi}/LAT \citetext{{\it Fermi Gamma-Ray Space Telescope}/Large Area Telescope, \citealp{2020ApJS..247...33A}} observations. In particular, it was reported that the $\gamma$-ray photon flux correlates with the compact radio flux density measured by the Very Long Baseline Array \citep[VLBA,][]{2009ApJ...696L..17K,2009ApJ...707L..56K}, and the jets of the LAT-detected AGNs have on average higher apparent speeds \citep{2009ApJ...696L..22L}, Doppler factors \citep{2010A&A...512A..24S}, fractional polarisations \citep{2010IJMPD..19..943H}, larger apparent opening angles and smaller viewing angles \citep{2009A&A...507L..33P}.

One of the most frequently given explanations of the observed $\gamma$-ray emission from AGNs is the inverse Compton scattering of soft photons by the jet relativistic electrons. Depending on where seed photons originate, several scenarios are proposed. Within the synchrotron self-Compton emission mechanism, the high-energy component of the blazar spectra is thought to be formed by up-scattered synchrotron photons \citep{1992ApJ...397L...5M, 1992NASCP3137..346M}. Other approaches consider external sources of photons, namely: thermal radiation from the accretion disk \citep{1993ApJ...416..458D}; scattered or reprocessed radiation inside the broad-line region \citetext{BLR; \citealp{1994ApJ...421..153S}} or infrared radiation produced by hot dust in the torus \citep{1995A&A...298..688W, 2000ApJ...545..107B}.

The knowledge of the $\gamma$-ray emission location could provide limits on the possible sources of the seed photons. At the most general level, the `close-dissipation' (sub-parsec, inside BLR) and `far-dissipation' ($\gtrsim$ 1~pc, outside BLR) scenarios are being discussed. In favour of the former, the main evidences are the GeV spectral break due to the photon-photon pair production \citep{2010ApJ...717L.118P} and short $\gamma$-ray variability time scales: few hours or even minutes \citep[e.g.][]{2010MNRAS.405L..94T, 2014ApJ...789..161N, 2016ApJ...824L..20A}. On the other hand, AGN studies in the radio band with single-dish or very long baseline interferometry (VLBI) monitoring programs indicate that the low- and high-energy emissions are co-spatial, supporting the far-dissipation scenario. \citet{2001ApJ...556..738J} found that the $\gamma$-ray flares detected by EGRET are likely to fall within 1$\sigma$ (typically 0.2~yr) uncertainties of the ejection epoch of the VLBI superluminal radio component. \citet{2003ApJ...590...95L} reported that the strongest $\gamma$-ray emission typically occurs during the rise or the peak of Mets\"ahovi 22 and 37~GHz radio flares associated with the ejection of a new VLBI component and obtained an average $\gamma$-ray emission region distance of about 5~pc from the core downstream the jet. Similar results were presented by \citet{2011A&A...532A.146L} on the basis of the LAT First Source Catalog data of 45 northern blazars. However, \citet{2010ApJ...722L...7P} found no significant correlation between the 15~GHz radio MOJAVE VLBA flux densities of downstream jet components and the LAT photon flux; instead, the correlation was strong for the VLBA core component, and the radio emission was lagging $\gamma$-rays. It was suggested that the time delay arises from the frequency dependence of the radio core radius $r_{\text{c}} \propto \nu^{-1}$, so that the $\gamma$-ray photons escape from the jet almost instantly, while it takes months for the propagating shock to reach the photosphere surface where the radio emission could be detected. For the cross-correlation analysis, \citet{2014MNRAS.441.1899F} used radio light curves obtained from the F-GAMMA programme and found that the time delays decrease from cm to mm/sub-mm bands, which is in good agreement with the theoretical prediction. \citet{2014MNRAS.445..428M} investigated the correlation between the \textit{Fermi}/LAT $\gamma$-ray flux and the 15~GHz OVRO radio flux density of 41 blazars and found that only three sources show correlations with larger than 2.25$\sigma$ significance, pointing out that observations over a longer time period are required to obtain higher significance. In the recent study, \citet{2019ApJ...877...39M} used different approaches to constrain the $\gamma$-ray emission region in the six \textit{Fermi}-bright flat-spectrum radio quasars (FSRQs), including the correlation analysis between $\gamma$-ray and radio light curves and a search for a spectral cutoff due to the pair production. All of them were consistent with rather large distances from the central engine (far beyond the BLR).

\textit{Fermi} has accumulated more than ten years of observational data of more than three thousand $\gamma$-ray bright blazars \citep{2020ApJS..247...33A}, providing great possibilities for studying the $\gamma$-ray emission observed in AGNs. In this study, we perform a cross-correlation analysis between the up-to-date \textit{Fermi}/LAT $\gamma$-ray flux and MOJAVE radio flux densities related to various VLBA components to address the following questions: (i) whether the $\gamma$-ray photons predominantly originate upstream of the 15~GHz VLBA core, (ii) whether the $\gamma$-ray emission zone is located within the BLR ($\lesssim 0.1$~pc) or beyond it at greater scales downstream the jet.

This paper is organised as follows. In \autoref{sec:sample} and \autoref{sec:data}, we describe the sample and the data, respectively. In \autoref{sec:method}, we describe the method used for the correlation analysis, and in \autoref{sec:delay} and \autoref{sec:localisation}, we present the analysis results. We summarise our main findings in \autoref{sec:conclutions}. 
Throughout this paper, we refer to the term `core' as the most compact opaque VLBI feature closest to the apparent base of the jet. We adopt a cosmology with $\Omega_m=0.27$, $\Omega_\Lambda=0.73$ and $H_0=71$~km~s$^{-1}$~Mpc$^{-1}$ \citep{Komatsu09}.

\section{Source sample}
\label{sec:sample}
    
    \begin{table}
    	\centering
    	\caption{Source sample. Columns are as follows: (1) B1950 name, (2) \textit{Fermi} 4FGL name, (3) number of radio epochs, (4) optical classification: Q -- quasar, B -- BL Lac, G -- radio galaxy, N -- narrow-line Seyfert 1 and U -- unknown spectral class, (5) redshift, (6) maximum apparent jet speed in units of the speed of light $c$ taken from \citet{MOJAVE_XVIII}. The full table consists of 331 sources and is available in its entirety in a machine-readable form.}
    	\label{t:sample}
    	\begin{tabular}{llccrr}
    		\hline
    		Source & \textit{Fermi} name & N & Opt. & $z$ & $\beta_{\text{app}}$ \\
    		(1) & (2) & (3) & (4) & (5) & (6) \\
    		\hline
    		0003$+$380 & J0005.9$+$3824 & 10 & Q & $0.229$ & $4.61\pm0.36$ \\
            0003$-$066 & J0006.3$-$0620 & 16 & B & $0.347$ & $7.31\pm0.33$ \\
            0006$+$061 & J0009.1$+$0628 & 5  & B & \dots & \dots \\
            0010$+$405 & J0013.6$+$4051 & 12 & Q & $0.256$ & $6.92\pm0.64$ \\
            0011$+$189 & J0014.1$+$1910 & 8  & B & $0.477$ & $4.54\pm0.46$ \\
            0015$-$054 & J0017.5$-$0514 & 8  & Q & $0.226$ & $0.72\pm0.28$ \\
            0016$+$731 & J0019.6$+$7327 & 7  & Q & $1.781$ & $7.48\pm0.50$ \\
            0019$+$058 & J0022.5$+$0608 & 7  & B & \dots & \dots \\
            0041$+$341 & J0043.8$+$3425 & 7  & Q & $0.966$ & $4.7\pm3.1$ \\
            0048$-$071 & J0051.1$-$0648 & 5  & Q & $1.975$ & $13.0\pm10.0$ \\
    		\hline
    	\end{tabular}
    \end{table}
    
    Our sample consists of 331 MOJAVE AGNs which have positionally associated $\gamma$-ray counterparts from the \textit{Fermi} LAT Fourth Source Catalog (4FGL-DR2). We do not consider a source if (i) there are less than five radio epochs available; (2) the galactic latitude $|b|<10^{\circ}$ (we checked that this cutoff does not affect the results of our study). The sample includes 198 objects which are a part of a complete set of 232 sources whose 15 GHz VLBA correlated flux density exceeds $1.5$~Jy at any epoch between 1994.0 and 2019.0 \citep{2019ApJ...874...43L}. The AGNs included in the current analysis are listed in \autoref{t:sample}; optical classes, redshifts $z$ and median apparent jet speeds $\beta_{\text{app}}$ for sources are taken from \citet{MOJAVE_XVIII} expect for the quasars 0414$-$189 and 1739$+$522 \citep{2019ApJ...874...43L}. In total, there are 194 quasars, 112 BL Lacs, 13 radio galaxies, six narrow-line Seyfert 1 galaxies and six sources with unknown spectral class in the sample. The redshifts are known for 281 (85\%) AGNs, with a median of 0.79. For a given source, $\beta_{\text{app}}$ is estimated as a maximum of the apparent speeds over all VLBA components. The apparent speeds obtained in that way are known for 261 (79\%) AGNs and have a median of $\sim$8.7~$c$. Finally, the most probable angle $\theta$ between the jet axis and the line of sight is estimated as \citep[Figure~2b]{2007ApJ...658..232C}: $\theta \sim 0.5 c \beta_{\text{app}}^{-1} \sim 0.06$~rad.

\section{Data}
\label{sec:data}
    
    \begin{figure*}
            \includegraphics[width=0.9\textwidth]{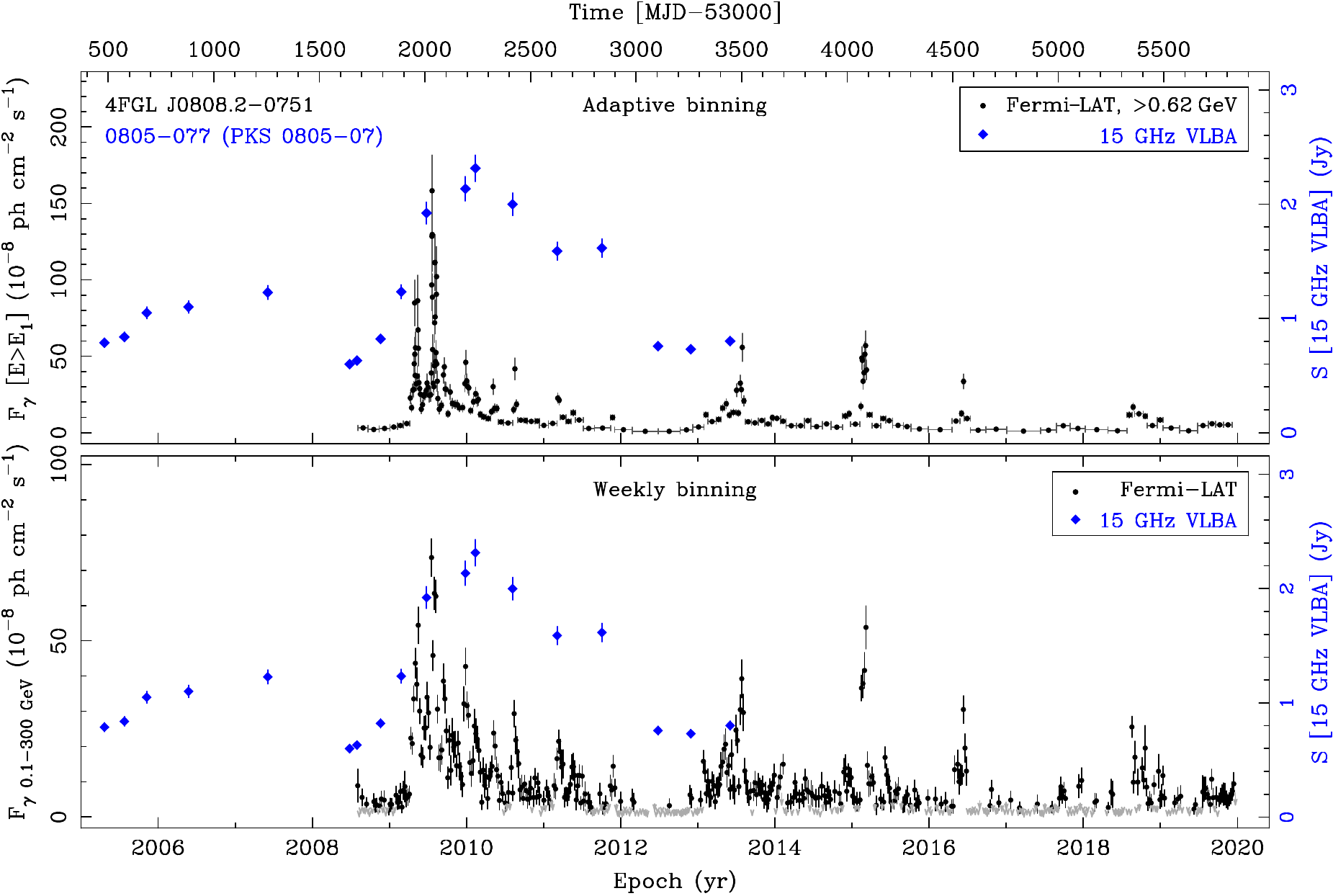}
            \caption{Adaptive (top) and weekly binned (bottom) $\gamma$-ray light curves for the quasar 0805$-$077.
            The grey arrows in the weekly binned light curves denote upper limits, with $\text{TS}<4$.
            The VLBA 15~GHz total flux density measurements are shown by blue.
            The light curves of all 331 sources are available in the online journal. 
            We plot and analyse the light curves after 2005 from the moment of the first VLBA or \textit{Fermi} data epoch.}
            \label{light_curve}
    \end{figure*}
    
	\subsection{Radio data}
	\label{sec:radio_data}
	
	The MOJAVE program \citep{2018ApJS..234...12L} is a long-term project aimed to investigate the parsec-scale properties of the brightest radio jets in the northern sky. In our study we used data of 4275 observations of 331 active galactic nuclei constituting our sample obtained with the VLBA at 15~GHz between 2005 January~6 and 2019 August~4 at 376 unique epochs \citep{MOJAVE_XVIII}. The radio flux densities of individual components were derived at each observing epoch by fitting a source brightness distribution with a limited number of circular or elliptical Gaussian components. The uncertainties of the VLBA 15~GHz flux densities are estimated to be about 5\% \citep{2002ApJ...568...99H,2005AJ....130.2473K}. We note that the radio data sets are unevenly sampled, and the median cadence varies significantly among the sources, ranging from 20 days to more than a year.
	
    \subsection{Gamma-ray data}
    \subsubsection{Weekly-binned light curves}
    
    To construct the weekly-binned $\gamma$-ray light curves for each source of our sample, we used the data obtained with the LAT \citep{LAT} onboard the {\it Fermi} $\gamma$-ray space telescope. In the analysis, we used the Fermitools software package version \textsc{v11r5p3} and followed a standard procedure recommended by the LAT team\footnote{\url{https://fermi.gsfc.nasa.gov/ssc/data/analysis/documentation/Pass8_usage.html}}. It includes two main parts, the binning and calculating the likelihood.
    
    We used the external script \textsc{make4FGLxml.py} to produce a source model by setting the 4FGL-DR2 source catalogue \citep{4FGL_DR2} based on the first ten years of LAT data, \textsc{gll\_iem\_v07.fits} as a background model of Galactic diffuse and isotropic emission based on the first eight years of LAT data (Pass 8 P8R3 source class events), \textsc{iso\_P8R3\_SOURCE\_V2\_v1.txt} for the isotropic spectral template and by selecting all sources within $20^\circ$ of the target. If an extended source was within the Region-of-Interest (ROI), at an angular distance less than $10^\circ$ from the target source, we considered its emission by running \textsc{gtdiffrsp} first. Otherwise, we treated the extended sources as point-like ones and used the  weekly photon files with the diffuse responses pre-computed. The source model file encapsulates information on a spectrum type, its corresponding characteristics and angular separation from the target for the considered sources. All spectral characteristics except the normalisation parameters of the sources within the ROI were kept constant during the likelihood analysis.
    
    Applying Pass 8, we made basic data cuts using the \textsc{gtselect} procedure, with (i) time range from 2008--08--04 through 2020--01--01, (ii) energy range 100~MeV -- 300~GeV, (iii) search centre at the source position and $\text{ROI} = 10^\circ$, from which we consider the source class photons, (iv) maximum zenith angle $90^\circ$ for energies above 100~MeV to filter out photons from Earth's limb as a strong source of background gamma rays. We also specified two hidden parameters: (i) $\text{event class} = 128$ as recommended for the point source analysis to only include events with a high probability of being photons and (ii) $\text{event type} = 3$ which includes all front+back converting events within all point spread functions and energy subclasses.
    
    Running \textsc{gtmktime}, we used the spacecraft file which contains the pointing and livetime history. We set $\text{DATA}\_\text{QUAL}=1$ and $\text{LAT}\_\text{CONFIG}=1$ to select the good time intervals (GTI), in which the satellite was properly working in the standard data taking mode, and the data were considered valid. The GTIs are used when calculating exposure. Next, we ran \textsc{gtltcube} to calculate the livetime as a function of `off-axis angle' and location on the sky for a specified observation time period using the size of a spatial grid of $1^\circ$. We also applied the zenith angle cut ($\text{zmax}=90^\circ$) to livetime calculation while running \textsc{gtltcube}, as currently recommended by the LAT team.
    
    Finally, \textsc{gtexpmap} was run to compute exposure maps setting a source region radius of $20^\circ$. This allows to take into account source and diffuse component emission beyond the ROI. This emission affects the sources within the ROI. This is due to the LAT point spread function, which is relatively broad at low energies ($3.5^\circ$ at 100~MeV). For exposure, the likelihood analysis and the calculation of the photon flux with the tool \textsc{gtlike} over each 7-day bin, we used the instrument response function \textsc{P8R3\_SOURCE\_V2}. The test statistic (TS) is defined as $\text{TS}=2\ln(L_1/L_0)$, where $L_1$ and $L_0$ are the likelihoods of the data with and without a point source at a considered position, respectively \cite[e.g.][]{Mattox96}. If a TS value of the 7-day bin was $<4$ (corresponding to about 2$\sigma$) or if the number of predicted photons $\text{N}_{\text{pr}}$ in that bin was $<10$, we calculated a 95\% upper limit of the photon flux \citep{Abdo11}. We skipped five weeks \textsc{w510--514} which embrace the period when the satellite was in a `safe hold' mode with instrument power off caused by the mechanical failure of the motor which rotates solar panels. 

	\subsubsection{Adaptive binning}
    \label{sec:adaptive}
    
    Following the method introduced by \cite{Lott12}, we also constructed adaptive binned $\gamma$-ray light curves for an integral flux within $0.1-300$~GeV, with constant relative flux uncertainty of $\sim20\%$ in each bin. The method assumes that the source energy spectrum is a single power-law function with photon index $\Gamma$ constant in time. The latter is well justified as the LAT-detected blazars manifest relatively moderate temporal variations in their spectra \citep{Abdo10_spec_prop}. The photon index was taken from the 4FGL-DR2 source catalogue. Binning was done using a normal time arrow and the source-dependent optimum lower energy cutoff $E_\text{min}$ at which the correlation between the photon flux and the spectral index is insignificant. The typical value of this parameter is around 1~GeV. The fluxes of bright sources in RoI were taken into account to improve the accuracy of the method. Compared with the fixed-binning approach, the adaptive binning method allows to extract more information encapsulated in the data and trace the bright flares in more detail deriving more accurate timing of the flare peaks, which is essential for the purposes of our study.
    
    The distribution of the bin widths for all the sources ranges from about 1~min for the strongest flares in the quasars 3C454.3 and 3C279 up to a few years for weak sources, with a median value of 11~days. The target relative uncertainty on the photon flux of 20\% holds for the fluxes above $\sim5\times10^{-8}$~ph~cm$^{-2}$~s$^{-1}$, with a tendency to higher errors for weaker flux bins comprising about 20\% of the bins' total number. This might occur due to the following reasons: (i) in case of a faint target source, modelling of the diffuse emission can be improper when calculating the bin widths, as the adaptive scripts do not do a full likelihood calculation while binning; (ii) the assumption of constant photon spectral index is not valid for the low-state bins; (iii) if the source is close to the Galactic plane, the assumption of a uniform level of diffuse emission over the RoI extent breaks down. These factors can contribute to the poor evaluation of the uncertainty. Their effects are amplified if a target source has a soft spectrum and is weak. For a few very weak sources, the algorithm has not yielded any single bin over the whole time period considered in our study.
    
    Both weekly and adaptive binning light curves in a tabular form are provided online.
    In \autoref{light_curve}, we present the $\gamma$-ray and radio light curves for the quasar 0805$-$077, as an example.
    
    \subsection{Correlation between the median radio flux density and gamma-ray flux}
    \label{sec:median_corr}
    
    \begin{figure}
        \centering
        \includegraphics[width=\linewidth]{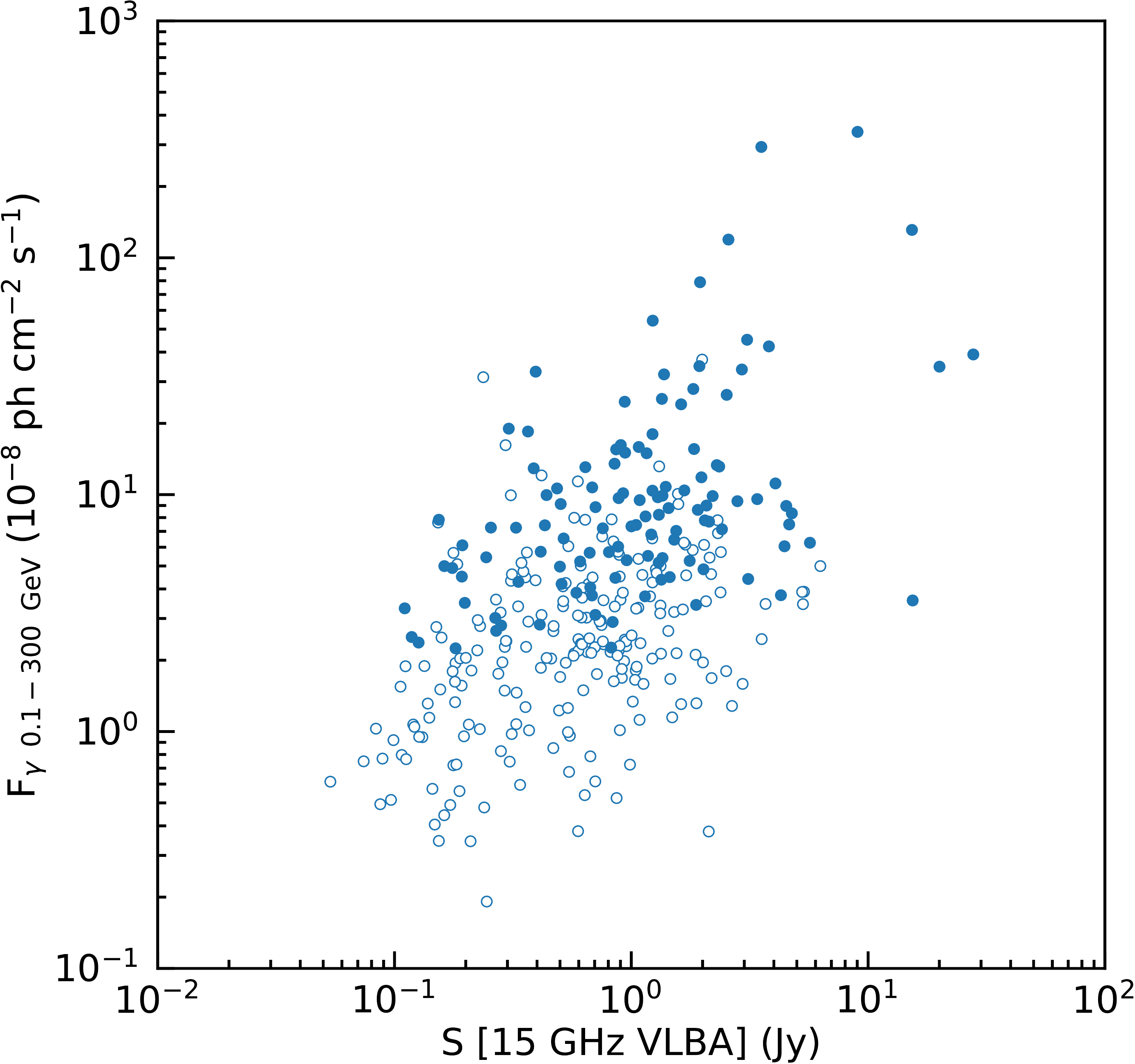}
        \caption{Median integrated \textit{Fermi} LAT 100~MeV -- 300~GeV photon flux taken from the adaptive binned light curves versus median total 15~GHz VLBA flux density for 331 sources from our sample defined in \autoref{sec:sample}. Each point represents a separate AGN. Data points treated as upper limits in the $\gamma$-ray band are denoted with unfilled markers. The BL Lac 0313$+$411 (4FGL~J0316.8$+$4120) has extremely low median $\gamma$-ray flux and is not included in the figure. }
        \label{fluxes}
    \end{figure}
    
    The non-parametric Kendall’s tau test performed between the median values of \textit{Fermi} LAT 100~MeV -- 300~GeV photon flux taken from the adaptive binned light curves and total 15~GHz VLBA flux density for all 331 sources in our sample shows a positive correlation coefficient 0.29 at a confidence level greater than 99.9\% (see \autoref{fluxes}) confirming early results of \citet{2009ApJ...696L..17K}. We investigated the effect of the upper limits in the $\gamma$-ray band by using \textsc{ASURV} Rev. 1.3 software package \citep{2014ascl.soft06001F}, which implements the methods of bivariate censored data analysis presented in \citet{1986ApJ...306..490I}. If more than a half of the data points in the $\gamma$-ray light curve were upper limits, then the median photon flux was also treated as an upper limit. The results of the survival analysis show that the correlation is present with a significance greater than 6$\sigma$. It should be noted that our sample is not flux-limited in either radio or $\gamma$-ray band, so the correlation is partly driven by the absence of sources in the bottom right and in the top left region in \autoref{fluxes}. However, this does not affect the position and the width of the peak of the stacked correlation curve (see \autoref{sec:stacked}) and might only change the absolute value of the correlation coefficient.

\section{Method of the correlation analysis}
\label{sec:method}
    
    \subsection{Individual correlation curves}
    
    To test for a possible time delay between the {\it Fermi}/LAT 100~MeV to 300~GeV $\gamma$-ray flux and the VLBA 15~GHz flux density we applied the z-transformed discrete correlation function \citetext{ZDCF; \citealp{10.1007/978-94-015-8941-3_14}} method to each source. The ZDCF is especially helpful when the data are sparse and unevenly sampled, which is the case for the radio light curves from our sample. We note that, in contrast to the ZDCF, another common approach --  the interpolation function \citep{1987ApJS...65....1G} -- implies that light curves vary smoothly and does not provide error estimates of the cross-correlation function. On the other hand, the discrete cross-correlation function method \citetext{DCF; \citealp{1988ApJ...333..646E}}, which is similar to the Pearson correlation coefficient, relies on the actual data only. However, it does not take into account that the sampling distribution of the correlation coefficient $r$ is often far from normal; thus, a more precise error estimation other than the standard deviation formula is required. Moreover, there is a significant difference between the ZDCF and the DCF approaches to the binning algorithm. The ZDCF binning algorithm shares the same idea with the adaptive binning in the sense that the bin width is varied to make sure that statistical significance is sufficiently high for each bin. In contrast, in the case of the DCF, this property might be violated due to a fixed bin width.
    
    The ZDCF binning algorithm consists of the following steps. First, for two given light curves, $r(t_\text{i})$ (radio) and $g(t_\text{j})$ ($\gamma$-ray; $t_\text{j}$ corresponds to the middle of j-th bin), time lags $\tau_\text{ij} = t_\text{i}-t_\text{j}$ are calculated and sorted in the ascending order. Note, a positive lag in our analysis corresponds to the $\gamma$-ray emission preceding the radio emission. Going from the median to the maximum time lag, all the pairs of fluxes are divided into bins of $n_\text{min}=11$ pairs; the procedure is repeated from the median down to the minimum time lag. To avoid interdependent pairs, a new pair is discarded if the same data point (either radio or $\gamma$-ray) is in the bin already. In addition, a new pair with the associated time lag $\tau''$ for which $|\tau''-\tau'|<\varepsilon$, where $\tau'$ is the time lag of the previous added pair and $\varepsilon$ is a small parameter, is allocated to the bin even if there are $n_\text{min}$ pairs already. This is done to prevent the artificial separation of close time-lags. While in the original algorithm $\varepsilon$ depends on the maximum time lag, we set it equal for all sources ($\varepsilon=3$~days). We checked that the choice of $n_\text{min}$, $\varepsilon$ and the first allocated bin does no affect the general features of the correlation curves, if the parameters vary within reasonable limits: $7 \lesssim n_\text{min} \lesssim 20$; $0.1\,\mathrm{d} \lesssim \varepsilon \lesssim 10\,\mathrm{d}$.

    \begin{figure}
        \centering
        \includegraphics[width=\linewidth]{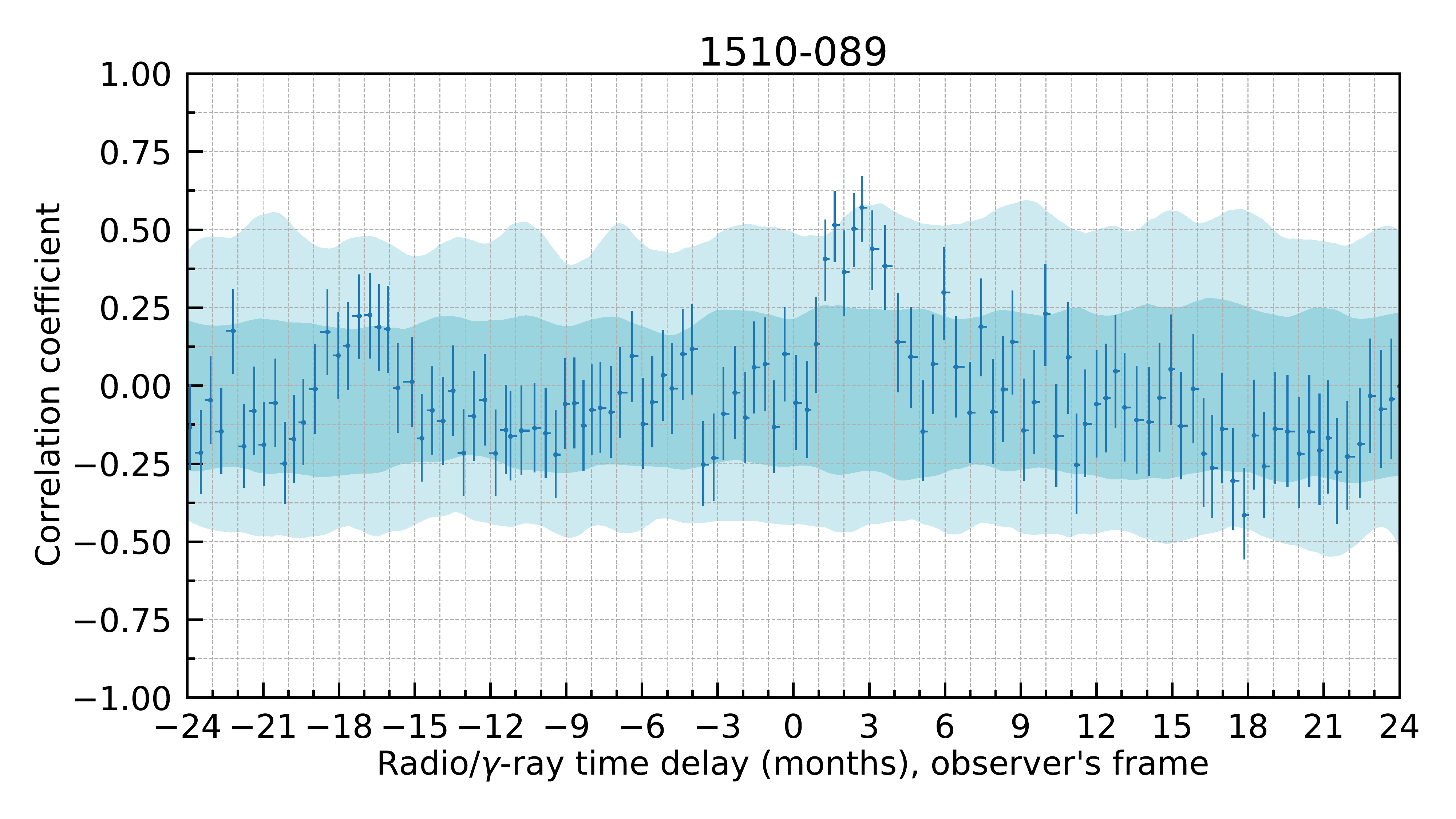}
        \includegraphics[width=\linewidth]{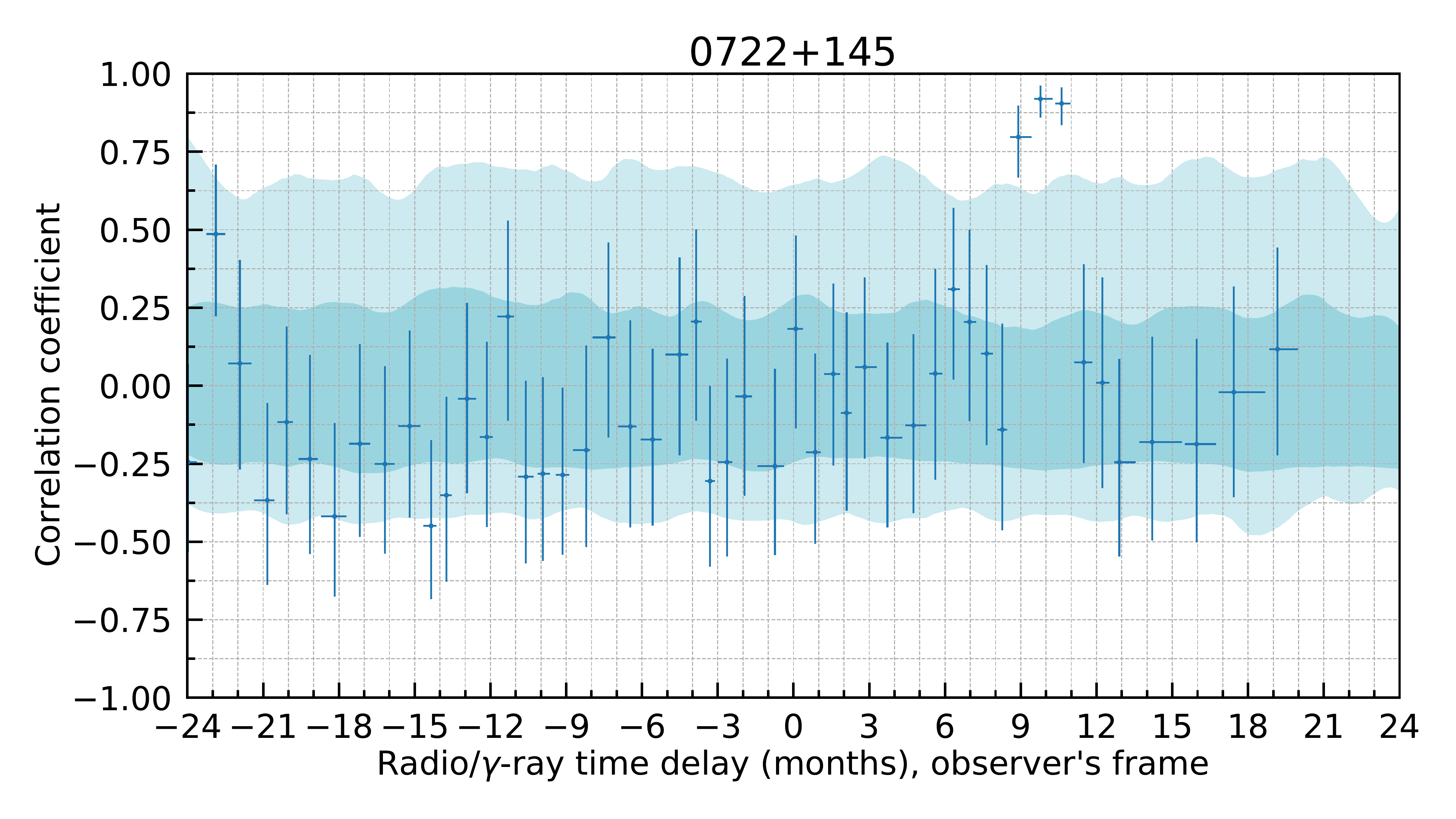}
        \includegraphics[width=\linewidth]{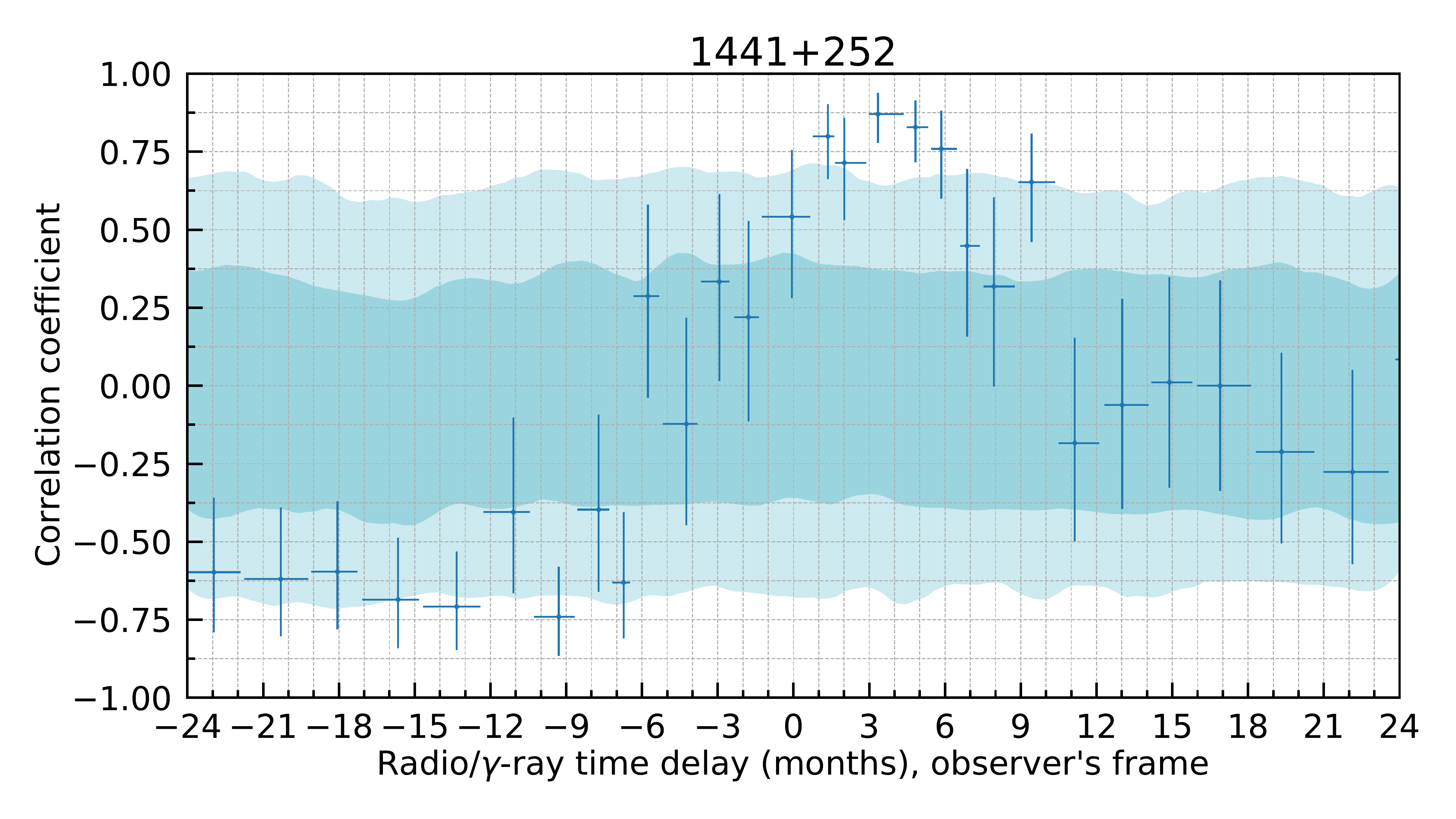}
        \caption{Individual ZDCF constructed for the quasars 1510$-$089, 0722$+$145 and 1441$+$252 with 55, 15 and 7 VLBA measurements, respectively. A positive time lag corresponds to the $\gamma$-ray emission leading the radio emission. The cyan and light-cyan areas indicate 68 and 95\% significance levels, respectively. The procedure of obtaining the significance levels is described in \autoref{sec:delay_ind}.}
        \label{zdcf}
    \end{figure}

	After the binning is done, the Pearson's $r$ correlation coefficient is calculated for each bin, and z-transformation is performed: $z=\tanh^{-1}{r}$. Then the mean $\overline{z}$ and the variance $s_z$ could be estimated (see equations 5 and 6 in \citealt{10.1007/978-94-015-8941-3_14}). The $\pm 1 \sigma$ intervals are calculated as follows:
	\begin{equation}
		\Delta r_{\pm} = |\tanh{(\overline{z} \pm s_{z})} - r| \,.
	\end{equation}
	To account for the flux errors, we performed Monte Carlo simulations $N_\text{mc}=300$~times, adding normally distributed errors to the data points and recalculating the ZDCF at each step. The average of $z$ is then used to derive $r$ and $\Delta r_\pm$. The resulting stacked correlation curves (see \autoref{sec:stacked}) tend to be smoother after this procedure.
	
    The ZDCF method requires at least $n_\text{min}=11$ data points for each light curve for a meaningful statistical interpretation. However, about 58\% of the sources from our sample have from five to ten radio epochs only. To include them in the analysis, we slightly eased restrictions so that a single data point could be presented twice in a bin at the cost of partial violation of pairs independence. Further, we refer to this sample as `total' in contrast to the reduced `robust' sample consisting of AGNs with $\geq 11$ radio epochs available. While we primarily draw conclusions from the robust sample, we also carry out the same analysis with the total sample to verify results on a larger number of sources.
    
\begin{figure}
        \centering
        \includegraphics[width=\linewidth]{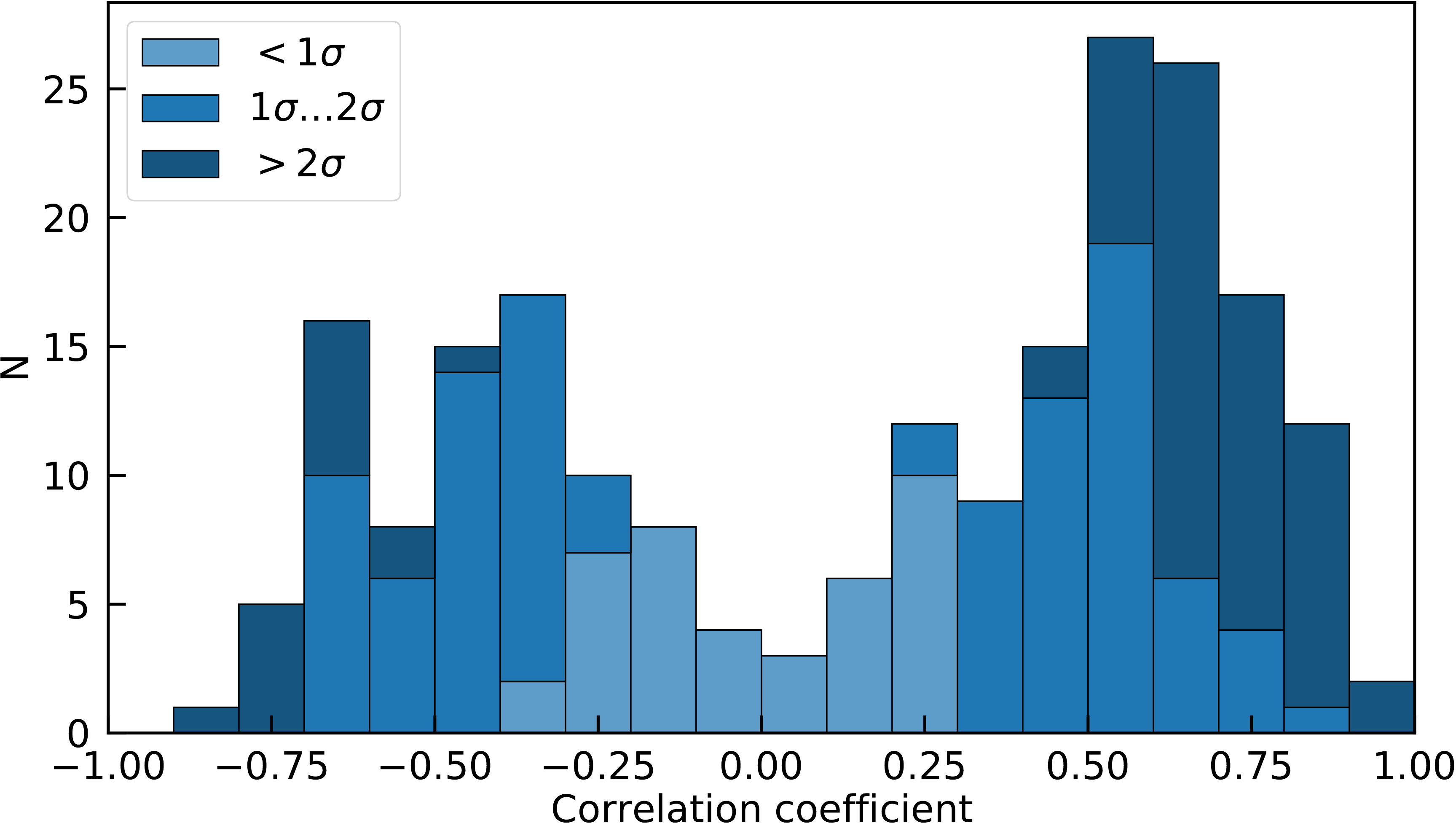}
        \caption{Distribution of the maximum absolute values of the correlation coefficients obtained from individual ZDCFs within the time delay interval of $[-1, 1]$~yr. The procedure of obtaining significance of the individual ZDCF data points is described in \autoref{sec:delay_ind}.}
        \label{zdcf_min_max}
    \end{figure}

    Examples of the individual ZDCFs constructed for the quasars 1510$-$089 (55 radio epochs), 0722$+$145 (15 radio epochs) and 1441$+$252 (7 radio epochs) are shown in \autoref{zdcf}. The calculations were done between the 15~GHz VLBA core flux densities and $\gamma$-ray 0.1--300~GeV photon fluxes taken from the adaptive binned light curves with parameters $\text{TS} > 7$ and $\text{N}_\text{pred} > 10$. An average ZDCF bin width increases with decreasing number of VLBA measurements available for a source, eventually reaching more than a month. However, in some cases there are still enough data to make statistically significant conclusions even for a source from the total sample with a small number of radio epochs (see \autoref{sec:delay_ind}). A distribution of the maximum absolute values of the correlation coefficients within the time delay interval of $[-1, 1]$~yr is shown in \autoref{zdcf_min_max}. After highlighting significance of these coefficients, it is evident that the positive correlation coefficients prevail over negative, confirming the overall correlation between the median radio flux density and $\gamma$-ray flux (\autoref{sec:median_corr}).
    
    \begin{figure*}
        \centering
        \includegraphics[width=0.49\linewidth]{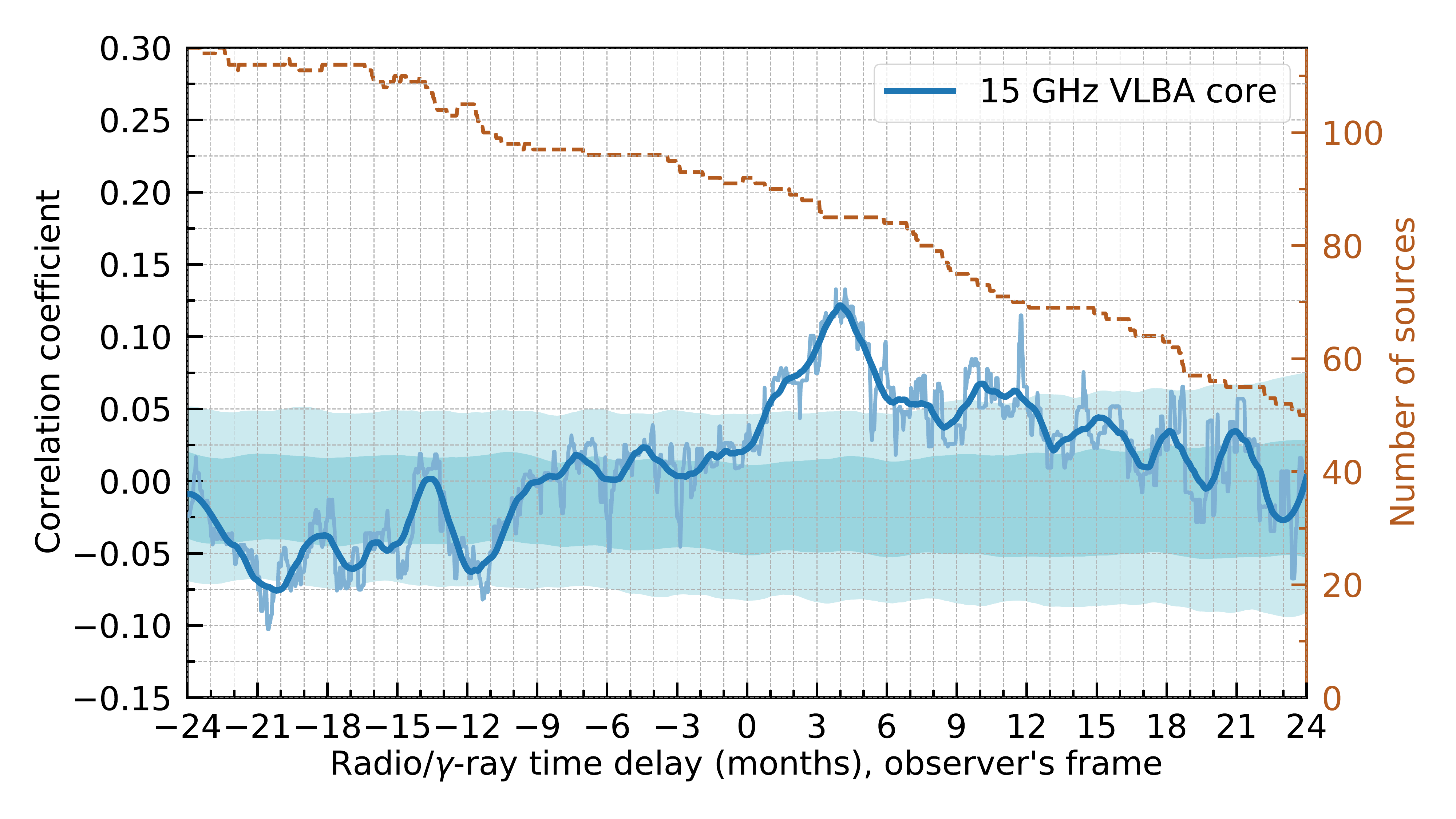}
        \includegraphics[width=0.49\linewidth]{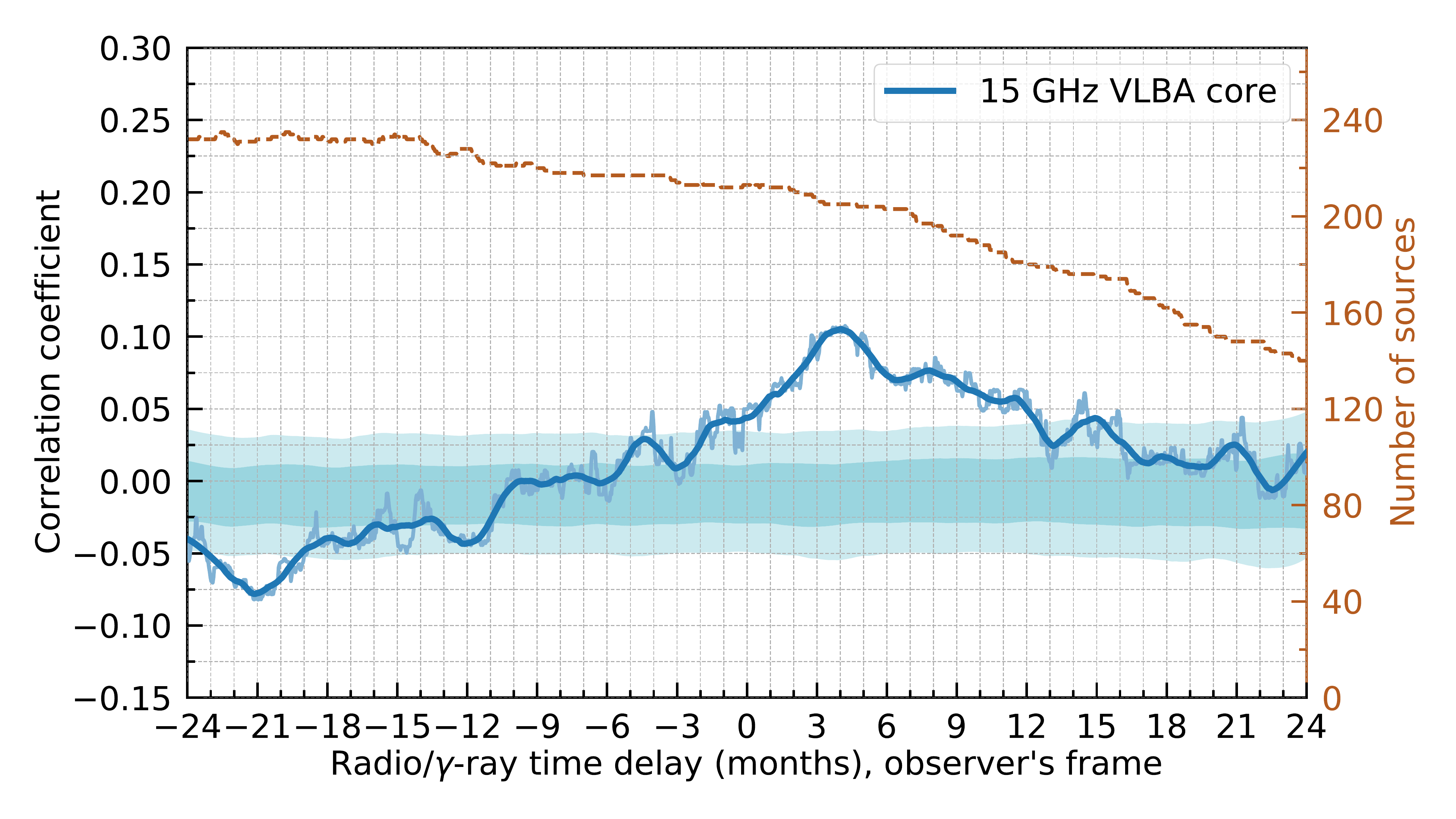}
        \includegraphics[width=0.49\linewidth]{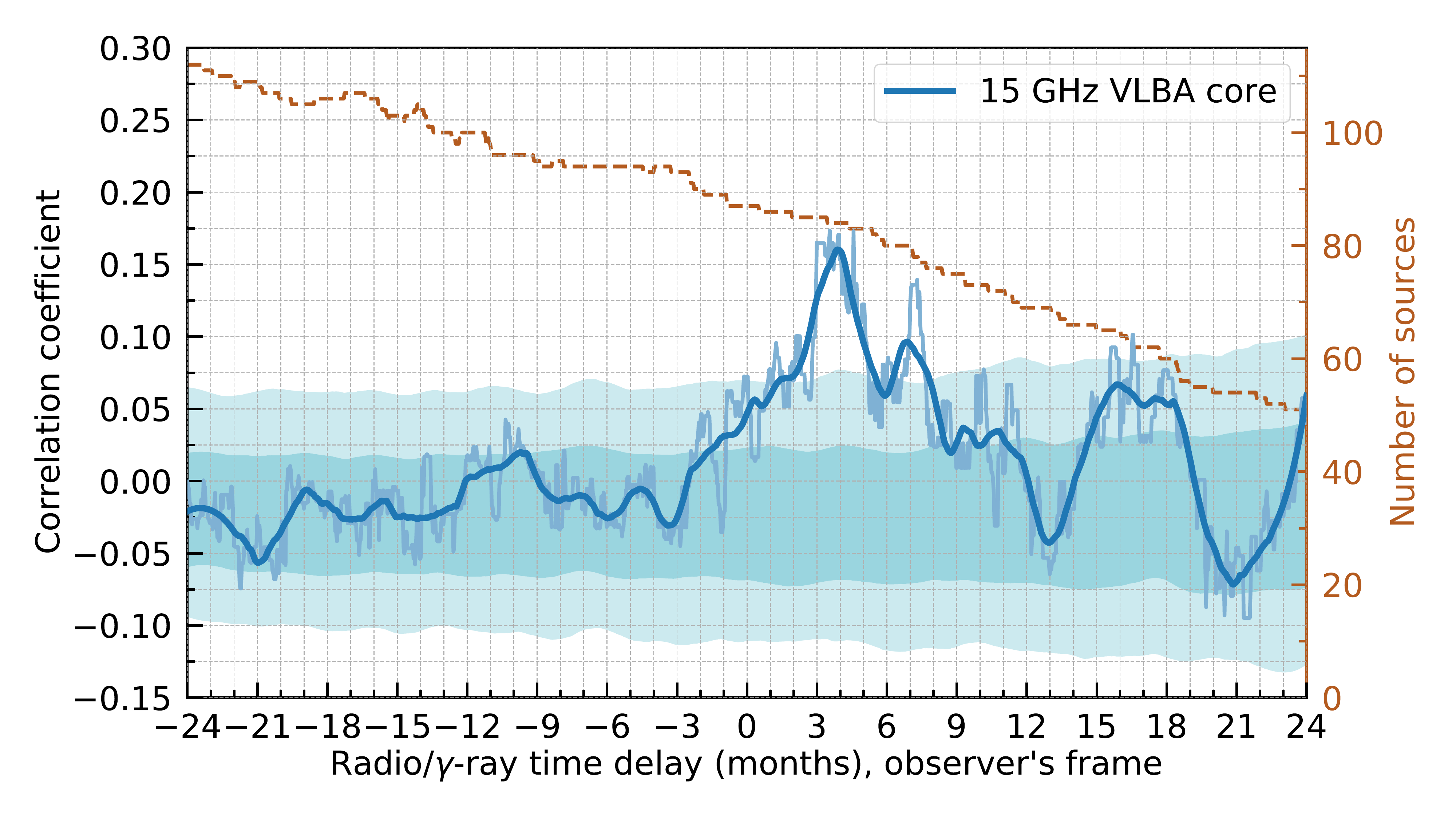}
        \includegraphics[width=0.49\linewidth]{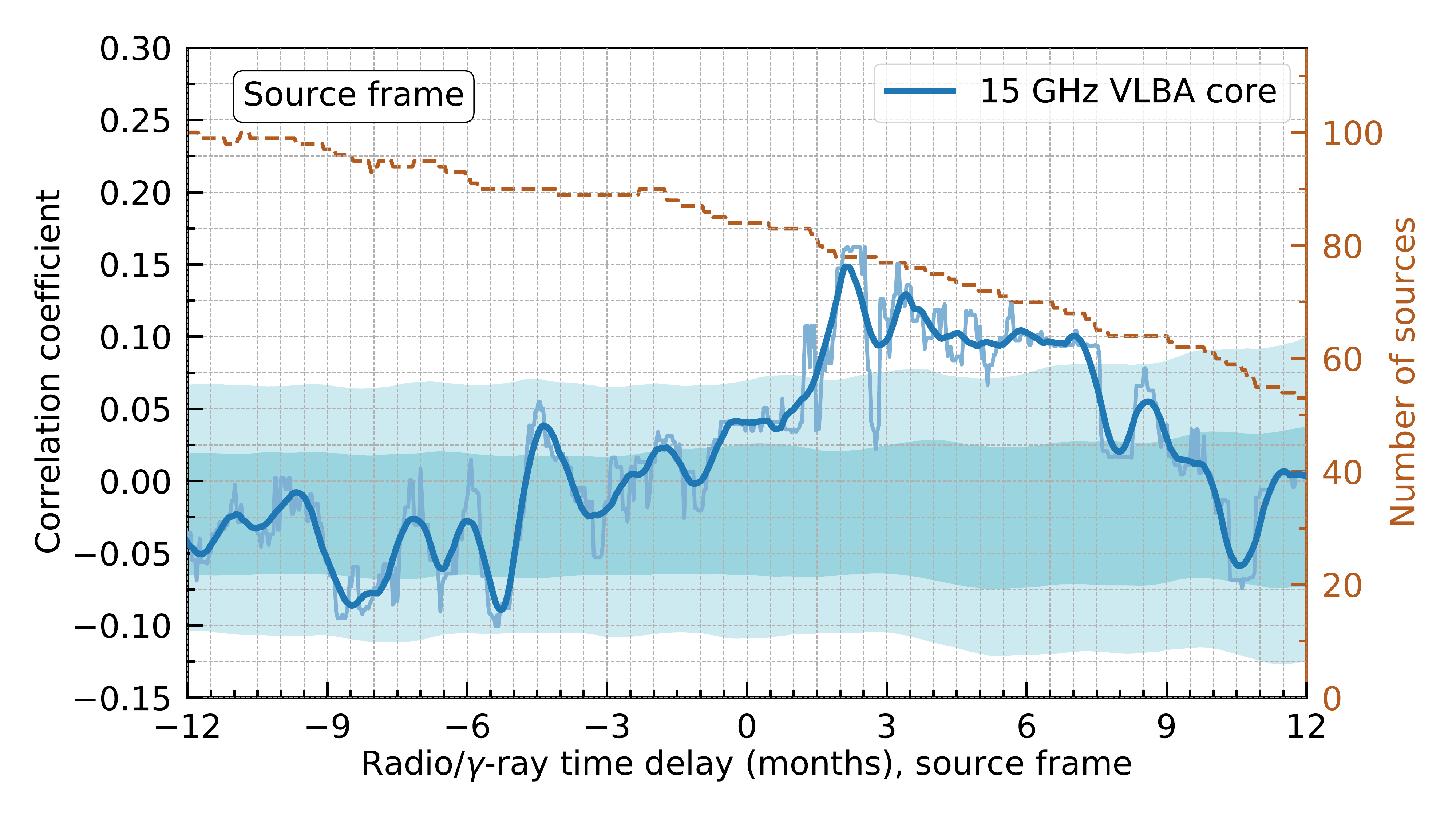}
        \caption{Top left: Stacked ZDCF representing the time lag dependent correlation between the weekly-binned $\gamma$-ray photon flux and the VLBA radio core flux density in the observer's frame (solid blue line). The original correlation curve before smoothing is shown in light-blue. A positive time lag corresponds to the $\gamma$-ray emission leading the radio emission. Only sources with more than 11 radio epochs are included in the analysis (`robust' sample). The cyan and light-cyan areas indicate 68\% and 95\% significance levels, respectively. The dashed brown line shows the number of sources which contribute to the median correlation coefficient, depending on the time lag. Top right: same curves, but the sources having from five to ten radio epochs are also included (`total' sample). Bottom left: robust sample; the adaptive instead of the weekly binned $\gamma$-ray light curves are used in the correlation analysis. Bottom right: same curves in the source frame.}
        \label{zdcf_stacked}
    \end{figure*}
    
    \subsection{Stacked correlation curves}
    \label{sec:stacked}
    
    After the individual correlation curves are done, the stacking correlation analysis is performed. First, we defined the values of the correlation function on the intervals between the ZDCF data points by assuming that the value is constant across the entire bin width. Then we calculated the median value of the correlation coefficients among all sources at the time delays which cover the considered time interval with a step of one day. It was verified that a variation of this parameter up to $\sim$30~days does not affect the main properties of the stacking curve (i.e. the maximum correlation coefficient and the width of the correlation peak). We also applied a Savitzky-Golay filter \citep{doi:10.1021/ac60214a047} with a 3-months window length in the observer's frame and 1.5-month in the source frame as well as 3$^{\text{rd}}$ order of the fitting polynomial to smooth the curve. To determine the robustness of the result, the null-level statistics was estimated as follows. We randomly shuffled the $\gamma$-ray light curves between the sources, keeping the radio data the same and recalculated the stacking correlation coefficients. This procedure was repeated 1000 times, and the 68\% and 95\% confidence intervals were defined as the intervals covering 16th-84th and 2.5th-97.5th percentiles of the correlation coefficient distribution, respectively.

\section{Radio/gamma-ray time delay}
\label{sec:delay}
    
    \subsection{Results of the stacking correlation analysis}

    We applied the stacking correlation analysis method to the 15~GHz VLBA core flux densities and weekly-binned $\gamma$-ray 0.1--300~GeV photon fluxes. For the $\gamma$-ray data we set test statistics $\text{TS} > 7$ and a number of predicted photons $\text{N}_\text{pred} > 10$ to avoid poor quality values in general and upper limits in particular. Even though the choice of the cutoff for the weekly binned light curves has a significant impact on the data points we use, since about one-half of the bins are discarded if $\text{TS} > 4$, the shape of the correlation curve and the time delay of the peak remain almost the same. The results for the robust sample are shown in \autoref{zdcf_stacked} (top left panel). A typical delay corresponding to the highest correlation coefficient (0.125) is 3-5~months in the observer's frame and delays, for which the correlation curve lies above the 95\% significance level, ranges from 1 to 8~months (no $>2\sigma$ significant negative lags are found within the considered time interval of $[-2, 2]$~yr). This interval of time delays is in a good agreement with our early results taken on a smaller sample \citep{2010ApJ...722L...7P}. The number of sources used for the stacking analysis decreases towards positive time delays, which is the sampling effect: the available radio data starts before the beginning of \textit{Fermi} observations and finishes half a year prior to the end of the $\gamma$-ray light curves. If the total sample is used, the position of the peak does not change, but it becomes smoother and at the same time the width of the confidence intervals decreases (\autoref{zdcf_stacked}, top right panel). We also repeated the analysis with sources which comprise a dominant part of the statistically complete flux-density-limited ($>1.5$~Jy) sample \citep{2019ApJ...874...43L} and found no significance difference between stacked correlation curves.
    
    \subsection{Weekly versus adaptive binning}
    
    If the $\gamma$-ray light curves with adaptive binning are considered, the stacked ZDCF peak associated with the core component becomes more pronounced, increasing from 0.125 to 0.17 in the observer's frame (\autoref{zdcf_stacked}, bottom left panel). This confirms our expectations that the adaptive binning method of light curves construction allows for better identification of rapid variability of the $\gamma$-ray emission. Short and strong flares are blurred in the weekly binned light curves, thus the correlation is established less accurately. In addition, the correlation results are almost insensitive to the TS and $\text{N}_{\text{pred}}$ cutoffs, which also argues in favour of the adaptive binning method.
    
    We also considered the brightest half of the sample based on the median $\gamma$-ray flux taken from the adaptive binned light curves and found that the correlation coefficient of the peak further increases to more than 0.2. The same holds true for the sources with apparent speeds greater than a sample median due to a present correlation between the fluxes and the apparent speeds. This difference indicates that the $\gamma$-ray flares of the brighter sources have better time resolution provided by the adaptive binning technique.

    \subsection{Source frame time delay analysis}
    
    \begin{figure}
        \centering
        \includegraphics[width=\linewidth]{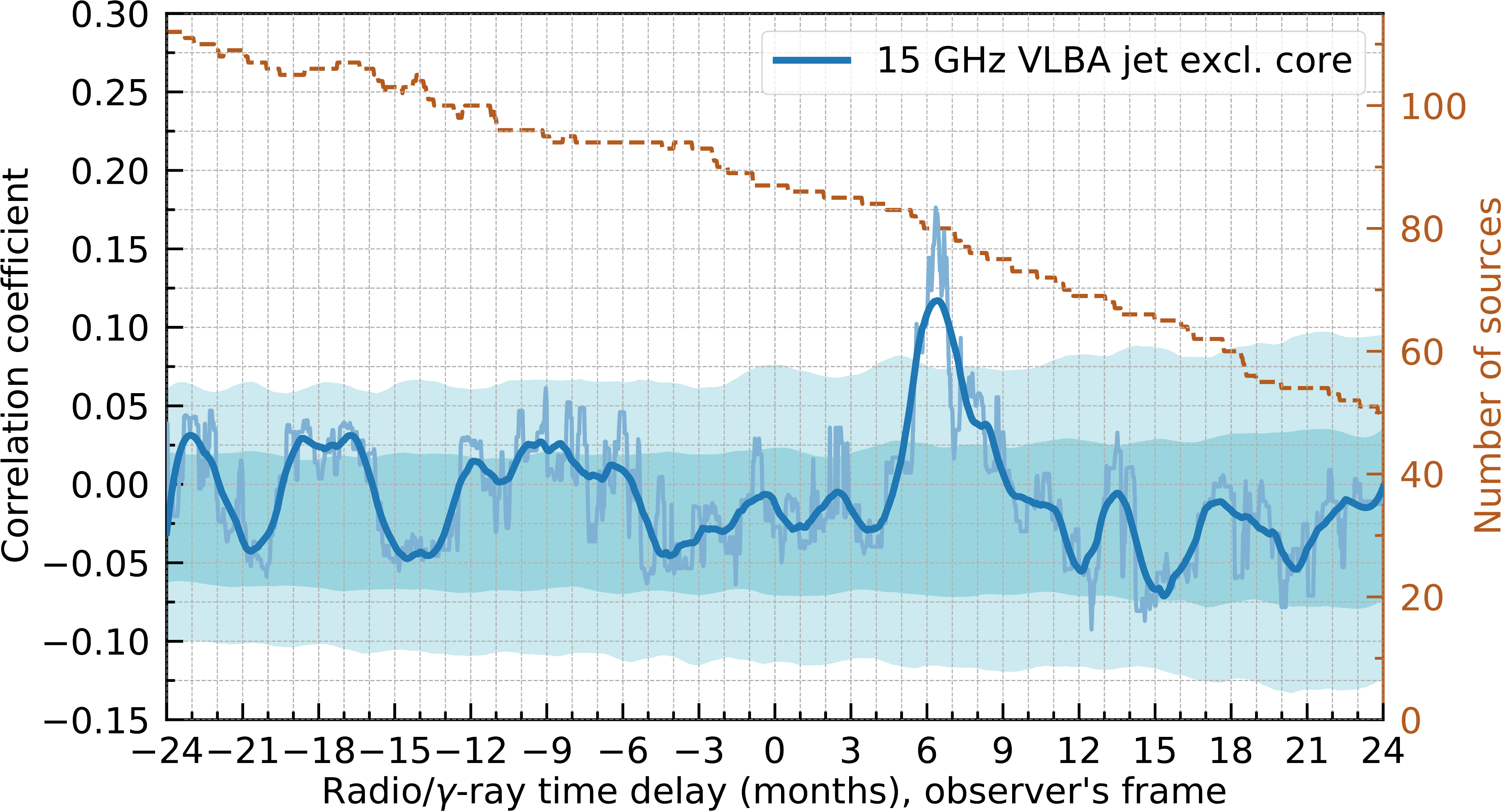}
        \caption{Same curves as on the bottom left panel in \autoref{zdcf_stacked}, but using the flux densities of the downstream VLBA components for the stacking correlation analysis. A positive time lag corresponds to the $\gamma$-ray emission leading the radio emission.}
        \label{downstream_comp}
    \end{figure}
    \begin{figure*}
        \centering
        \includegraphics[width=0.49\linewidth]{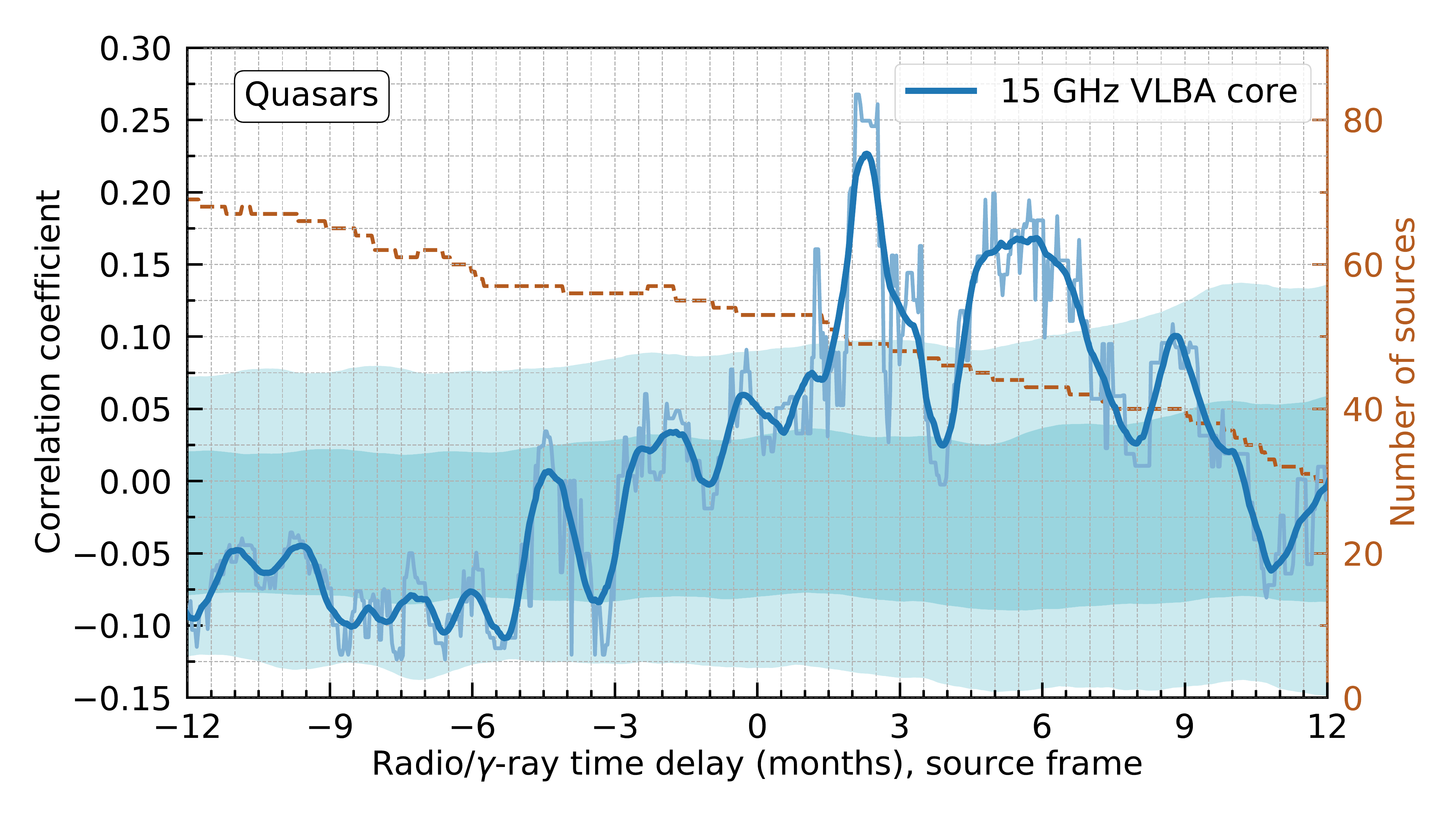}
        \includegraphics[width=0.482\linewidth]{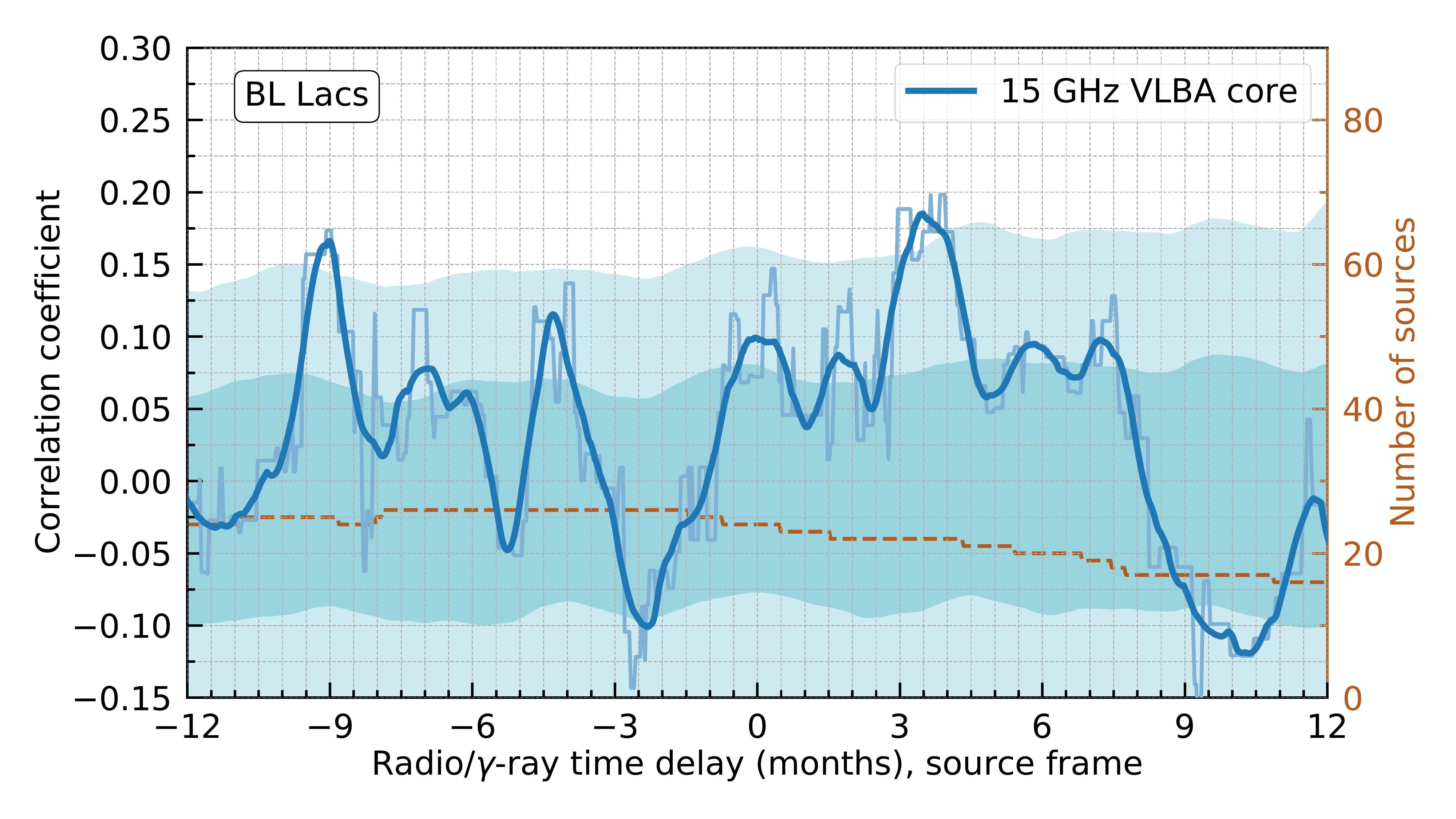} 
        \caption{Same curves as on the bottom right panel in \autoref{zdcf_stacked}, but separately for the quasars (left) and the BL Lacs (right). A positive time lag corresponds to the $\gamma$-ray emission leading the radio emission.}
        \label{quasars_bllacs}
    \end{figure*}
    
	In addition we performed the stacking analysis in the source frame by correcting radio/$\gamma$-ray time delays by a factor of $(1+z)^{-1}$ for each source in accordance with its redshift (the time interval was kept the same). We did not apply the K-corrections to the fluxes because it does not influence the correlation coefficients as the Pearson's $r$ is scale invariant. The sources with unknown redshifts were excluded from the analysis. Since for our sample the median $z \approx 0.75$, the peak of the correlation curve is expected to be at about two times smaller time lag. This can be clearly seen on the bottom right panel in \autoref{zdcf_stacked}: a typical delay decreases to 2-3~months in the source frame. The peak is still ensured by about one half of the robust sample, and its absolute value remains almost unchanged, exceeding the 95\% significance level. The correlation peak found by \cite{2010ApJ...722L...7P} was at about 1.2~months. The difference can be driven by a number of reasons: our current sample is twice as large, the light curves we use are an order of magnitude longer, and we apply a different method of time delay assessment.
	
    \subsection{Downstream jet components}
    \label{sec:downstream}
    
    In general, the radio flux density of the core component far exceeds flux densities of other components in parsec-scale radio jets of Doppler-boosted blazars \citep[e.g.][the procedure of obtaining the flux densities of individual components is described in \autoref{sec:radio_data}]{2005AJ....130.2473K,RDV_PK12}. Note, downstream components usually include newly ejected features as well as the remnants of the old ones, and therefore their total flux is variable. We performed the ZDCF stacking analysis with total flux density of downstream parsec-scale components only and found that for different reasonable TS and $\text{N}_{\text{pred}}$ cutoff values, the correlation coefficient remains within 68\% significance level for time delays which are typical for the core component, 5~months and less. Instead, the correlation curve tends to have a peak at time delays of 5-9~months. Indeed, the analysis conducted with the adaptive light curves displays 95\% significant peak at 6-7~months (\autoref{downstream_comp}).
    
    \citet{2002A&A...394..851S} have shown that almost for every new ejected VLBA component, there is a coincident total flux density flare. However, it was also reported that during the outbursts, it is typically the \textit{core} that shows considerable flux density variability. We suggest that the rise of the correlation with external components corresponds to the radio flares interpreted as plasma disturbances, which initially passed through the core region and then reached the downstream VLBA components. Therefore, the primary correlation is that of between the $\gamma$-ray flux and the radio core flux density. The peak in \autoref{downstream_comp} is a consequence of the flux density evolution from the core to the downstream components.
    
    \subsection{Quasars versus BL Lacs}
    
    \autoref{quasars_bllacs} demonstrates that in the source frame the range of typical time delays is similar between the quasars and BL Lacs, but the correlation is more significant for the former. One of the possible explanations is that the number of quasars included in our sample is significantly greater than the number of BL Lacs (194 vs 112). In addition, the quasars are brighter in both the radio and $\gamma$-ray bands, therefore their individual ZDCFs have lower errors. We note that the stacked correlation curve of BL Lacs has a correlation peak at $-$9~months and a sub-peak at $-$5~months which is inconsistent with the interpretation given in \autoref{sec:downstream}. We repeated the stacking analysis with the BL Lacs using the total sample and found that these peaks no longer exceed the 68\% significance level, while the peak at 4~months remains significant. Considering that the effective number of BL Lacs available to calculate the median correlation coefficient is extremely low (typically $\sim$25 or less), we suggest that the peaks at the negative time delays do not reflect any real physical phenomena.

\section{Gamma-ray emission localisation}
\label{sec:localisation}
    \subsection{Radio/gamma-ray delay and core opacity}
    
    The stacking correlation analysis revealed that the $\gamma$-ray emission is likely to be produced between the nucleus and the 15~GHz VLBA core. The question remains, however: is the source of the seed photons located inside or outside the BLR? To address this problem, we could estimate the distance between the regions of the radio and $\gamma$-ray emission production, taking the radio core opacity as the main source of the time lag \citep[e.g.][]{2012A&A...545A.113P,2019MNRAS.486..430K}.
    
    The time delay obtained from the ZDCF is a measure of the time interval between the \textit{peaks} of the radio and $\gamma$-ray flares. Assuming that the peak of the radio core flare coincides with the moment when the plasma disturbance passes through the VLBI core position, we have the following equation:
    \begin{equation}
        \Delta r = \frac{\beta_\text{app} \Delta t_\text{obs}}{(1+z)\sin{\theta}} \,,
        \label{rg_dist}
    \end{equation}
    where $\Delta r = r_\text{c} - r_\gamma$ is the distance between the radio core ($r_\text{c}$) and the place of the $\gamma$-ray emission production ($r_\gamma$), $\beta_\text{app}$ is the apparent jet speed, $\theta$ is the jet viewing angle, $z$ is the redshift and $\Delta t_\text{obs}$ is the time delay in the observer's frame. Substituting typical parameters for our sample (see \autoref{sec:sample}) into this equation results in a distance of several parsecs. On the other hand, the distance between the core as an apparent origin of the jet and the true jet apex can be estimated as follows:
    \begin{equation}
        r_\text{c} \approx \frac{\Omega}{\nu\sin{\theta}} \,,
        \label{rc_dist}
    \end{equation}
    where $\nu$ is the observed frequency in GHz and $\Omega$ is the core shift measure. The latter is defined in \citet{1998A&A...330...79L} as:
    \begin{equation}
        \Omega = 4.85 \cdot 10^{-9} \frac{\Delta r_\text{mas} D_\text{L}}{(1+z)^2} \frac{1}{\nu_1^{-1/k_\text{r}} - \nu_2^{-1/k_\text{r}}}\, \text{pc} \cdot \text{GHz}^{1/k_\text{r}} \,,
    \end{equation}
    where $\Delta r_\text{mas}$ is the core shift in milliarcseconds, $D_\text{L}$ is the luminosity distance in parsecs and $k_\text{r}$ is the power index (we assume the equipartition between the particle and magnetic field energy density: $k_\text{r} = 1$, see \citealp{2011A&A...532A..38S}). The core shift measurements yield $r_\text{c}$ values of the order of several parsecs \citep[e.g.][]{2019MNRAS.485.1822P} -- similar to the discussed above $\Delta r$. Thus, the stacking correlation analysis does not allow us distinguishing between the two scenarios of the high-energy emission production. A calculation of the time delays related to individual sources might clarify the situation.
    
    \subsection{Time delays of individual sources}
    \label{sec:delay_ind}
    
    To obtain the time delays and estimate their uncertainties for each source individually, we follow the likelihood method described in detail in \citet{2013arXiv1302.1508A}. Further in this section, we operate with the total sample and study the ZDCFs constructed in the observer's frame with the use of the adaptive-binned $\gamma$-ray light curves and the radio core flux densities. First, we specify the time range where to search the peak: $[-2, 9]$~months, which is the interval where the stacking correlation coefficient exceeds the 68\% significance level. This partly prevents from detecting the peaks associated with physically unrelated flares. All sources having less than five ZDCF data points in the selected range are excluded from further consideration. The likelihood of a data point \textit{i} being ZDCF peak is defined as a product of probabilities that \textit{i} is larger than any other point. Within this approximation the maximum likelihood estimate $\Delta t_\text{obs}$ always coincides with the position of the ZDCF maximum. After linearly interpolating between the points of the likelihood function, the uncertainty of the peak $\Delta t_\text{obs}^{\pm}$ is estimated by defining the fiducial interval which covers 68\% of the area around the maximum. We exclude from the analysis all time delays with extremely high uncertainties ($>150$~days).
    
    \begin{figure}
        \centering
        \includegraphics[width=\linewidth]{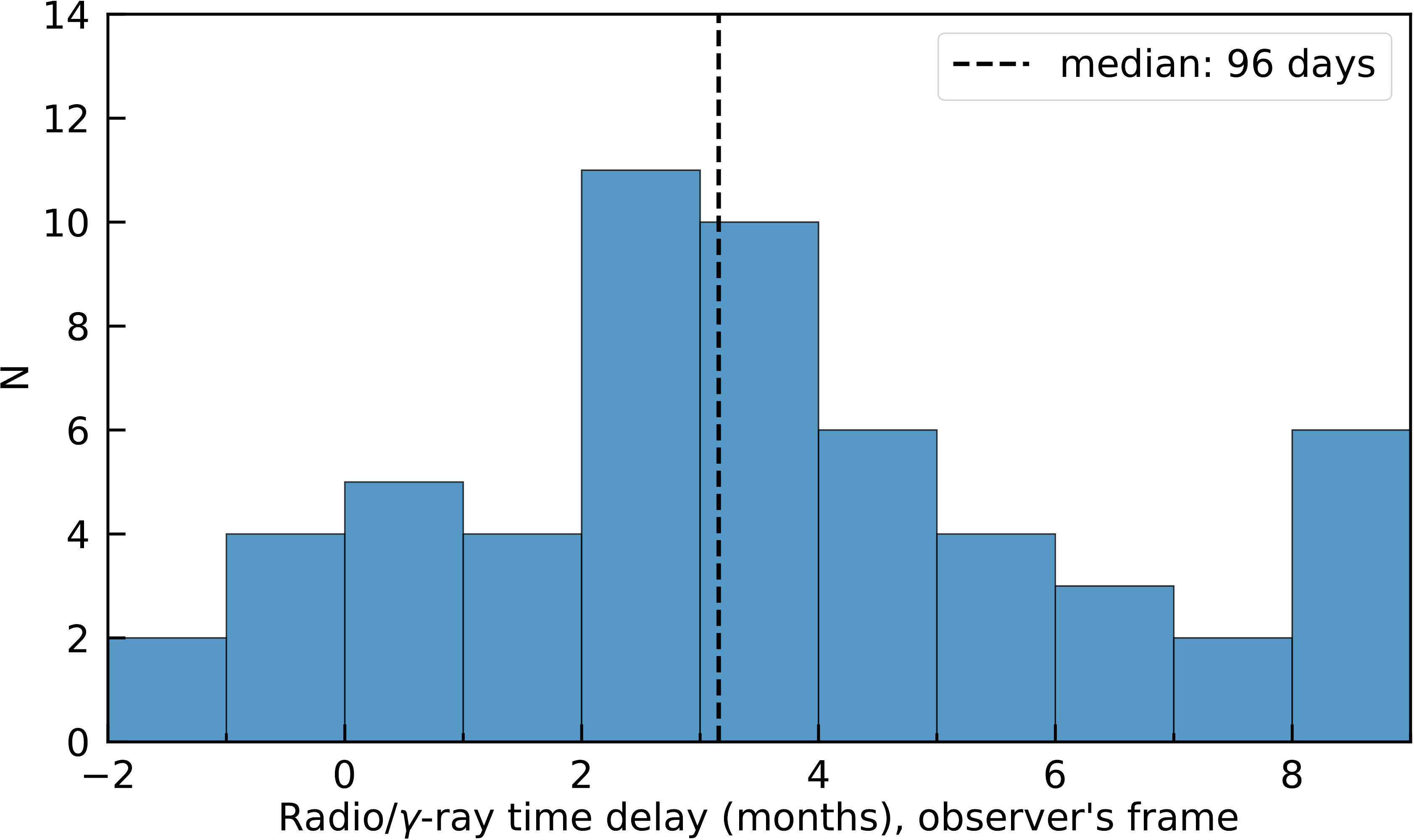}
        \includegraphics[width=\linewidth]{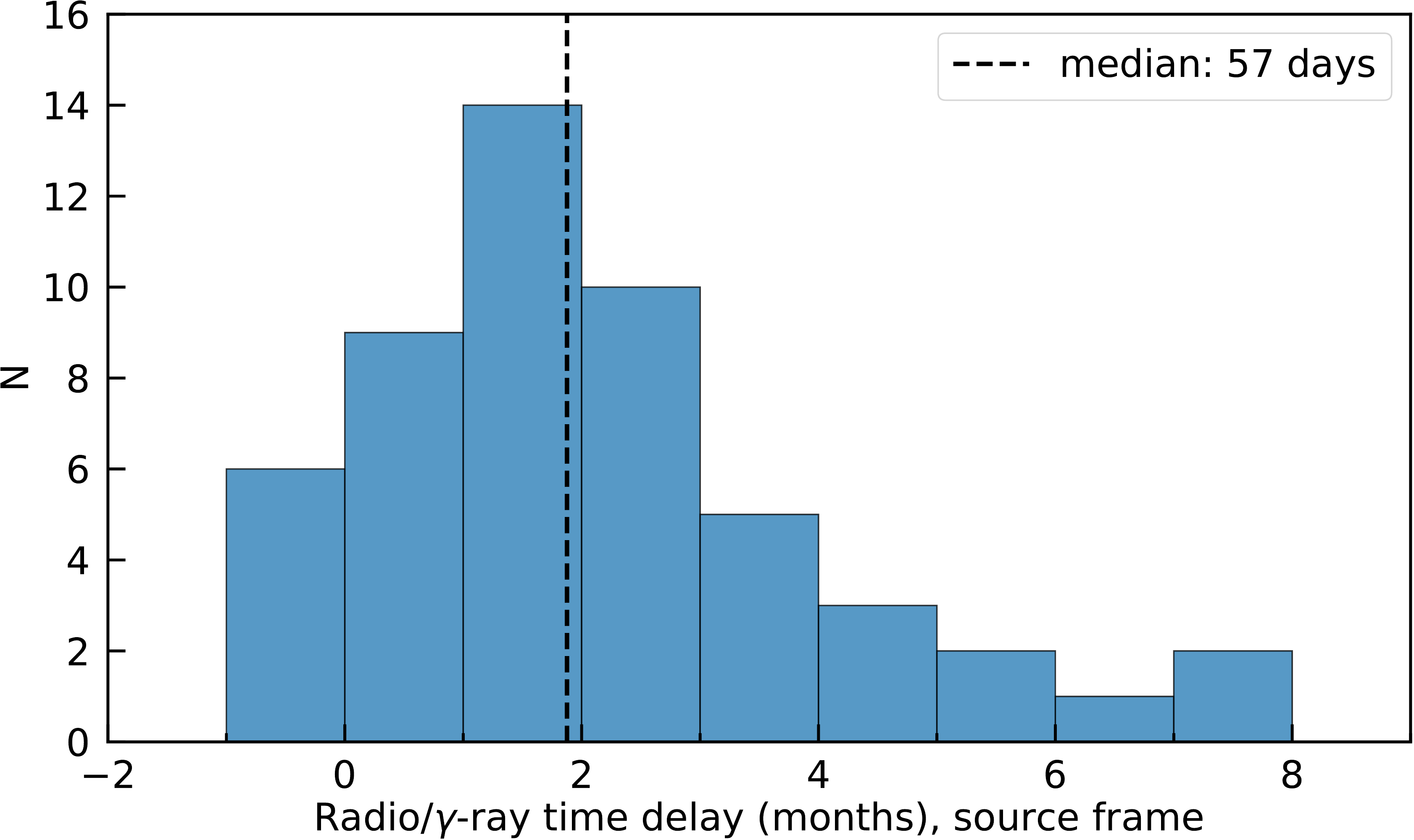}
        \caption{Top: distribution of radio/$\gamma$-ray time delays obtained from the individual ZDCFs (observer's frame) for 57 sources with a median time lag of 96 days. Bottom: the same distribution in the source frame (52 sources with known redshifts).}
        \label{lags_hist}
    \end{figure}

    \begin{figure}
        \centering
        \includegraphics[width=\linewidth]{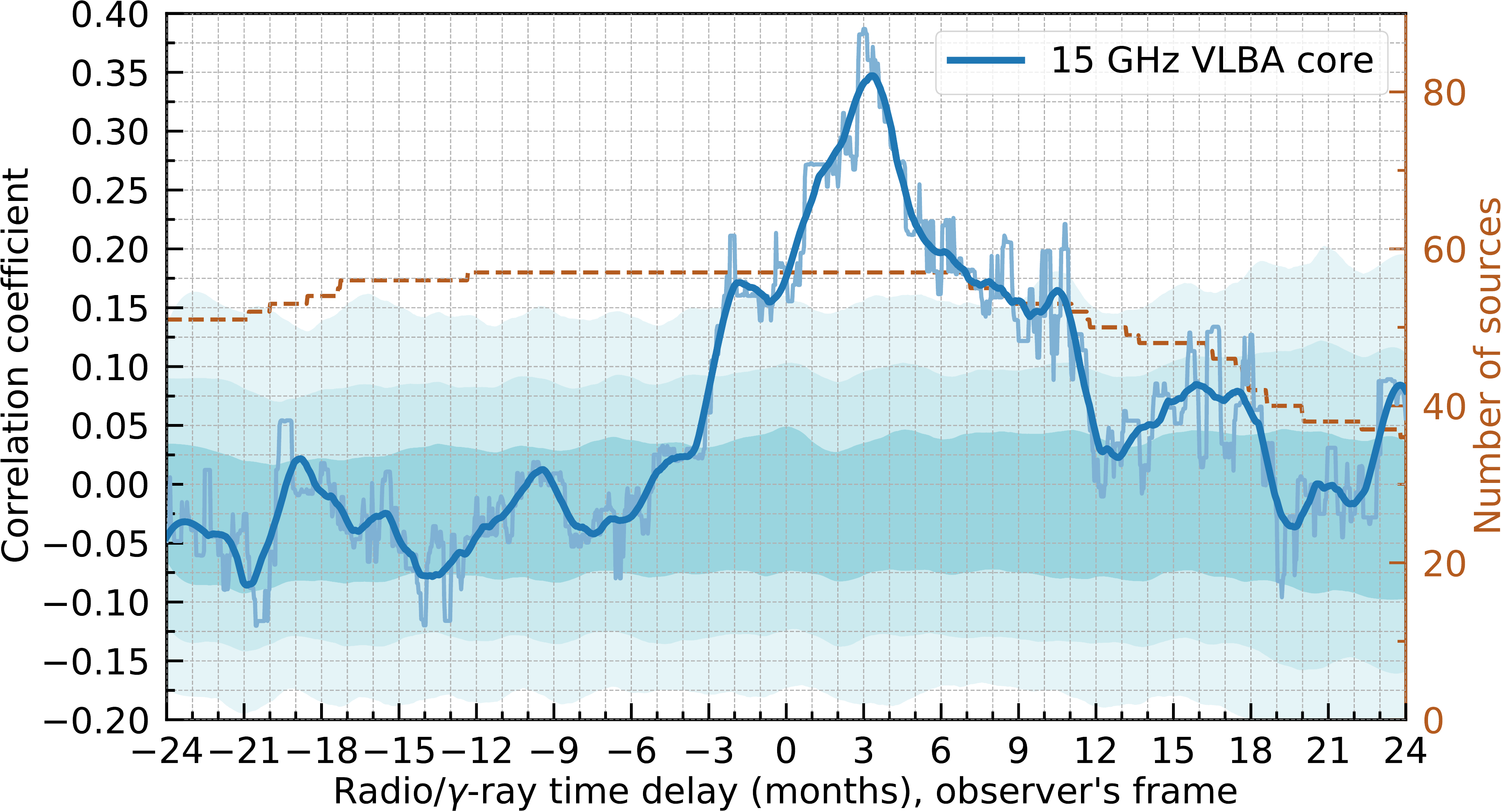}
        \caption{Same curves as on the bottom left panel in \autoref{zdcf_stacked}, but the sample is limited to the sources with the obtained individual radio/$\gamma$-ray time lags. A positive time lag corresponds to the $\gamma$-ray emission leading the radio emission. The lightest cyan area indicates 99.7\% significance level.}
        \label{sample57}
    \end{figure}
    
    The significance of the time lags was estimated in the same manner as the significance levels in the stacked correlation curves. We calculated the ZDCFs for each source for which the time lag was identified, keeping radio data the same and taking all $\gamma$-ray light curves from the total sample, one by one. It was found that the time lags of 40\% of the sources (29 out of 73) have more than 2$\sigma$ significance, which increases the number of significant correlations compared to \citet{2014MNRAS.445..428M}. However, there is also a considerable number of sources (22\%) whose ZDCF peaks have low ($<1\sigma$) significance and we exclude them from further consideration.
    
    The resulting distribution of the time lags calculated for 57 sources is shown in \autoref{lags_hist} (top panel); the numerical values and their uncertainties are presented in \autoref{t:time_lags}. A relatively small number of sources for which the time delays were identified ($\sim$1/5 of the total sample) is due to sparse radio data and, as a consequence, the lack of the ZDCF data points in the specified time range. The fact that the distribution has a median of 96 days and one-half of the individual time delays lies in the range from 2 to 5~months is consistent with the position of the peak in the stacked correlation curve (\autoref{zdcf_stacked}, left bottom panel). A similar consistency holds for the time lags in the source frame (\autoref{lags_hist}, bottom panel). We also performed the stacking correlation analysis with the sources with determined time delays and found that the peak becomes considerably more prominent, exceeding 0.35 level (\autoref{sample57}). We repeated the shuffling procedure from \autoref{sec:stacked} 3000~times to obtain the 99.7\% significance level and found that it is lower than the ZDCF peak. However, not only this part of the sample contributes to the stacking correlation coefficient. The stacking analysis conducted with the rest of the sources only also resulted in a peak at the similar time delays, but with lower significance, though.
    
    \begin{figure}
        \centering
        \includegraphics[width=\linewidth]{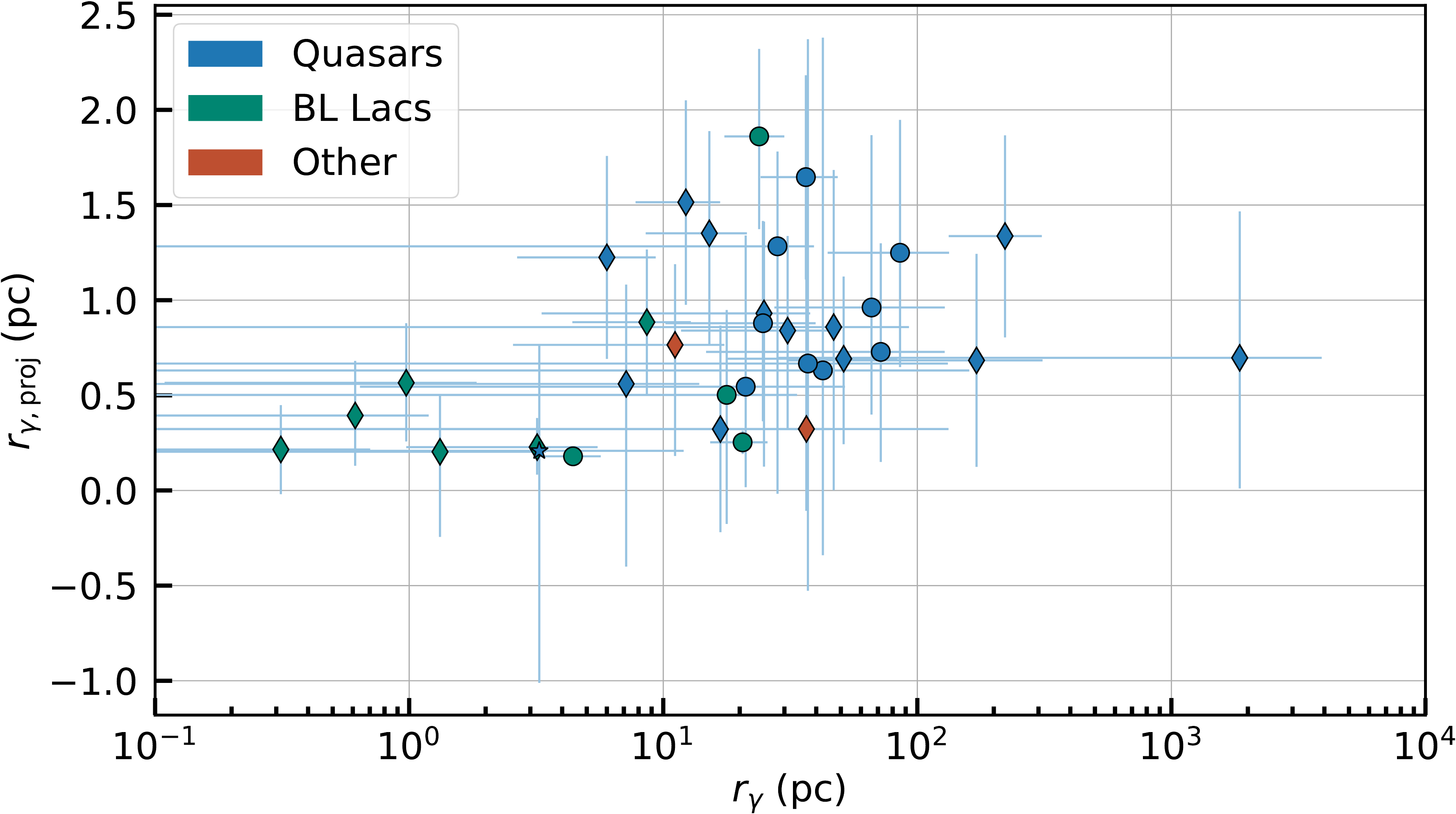}
        \caption{Distance between the central engine and the region of the $\gamma$-ray emission obtained from the individual ZDCFs and the core shifts. The horizontal axis represents the absolute distance; the vertical axis represents the distance projected onto the sky plane. The blue, green and orange data points refer to the quasars, BL Lacs and other source classes, respectively. The core shifts are taken from \citet{2012A&A...545A.113P} (circles) and Plavin et al.\ (in prep.; asterisks), or the median from \citet[]{2012A&A...545A.113P} is used (thin diamonds). Only positive distances are shown in the plot.}
        \label{rgamma_hist}
    \end{figure}
    
    \begin{table}
    \centering
    \caption{Derived radio/$\gamma$-ray time delays $\Delta t_{\text{sour}}$ together with 1$\sigma$ uncertainties: 57 sources, observer's frame.}
    \label{t:time_lags}
    \begin{tabular}{cl|cl}
    		\hline
    		Source & $\Delta t_\text{sour}$ & Source & $\Delta t_\text{sour}$ \\ 
    		name   & (days)                 & name   & (days)                 \\
    		\hline
    		0110$+$318 & $11_{-25}^{+70}$   & 1236$+$049 & $-19_{-17}^{+82}$  \\[0.5ex]
            0130$-$171 & $50_{-21}^{+94}$   & 1308$+$326 & $96_{-11}^{+16}$   \\[0.5ex]
            0141$+$268 & $68_{-64}^{+35}$   & 1329$-$049 & $200_{-121}^{+38}$ \\[0.5ex]
            0208$+$106 & $73_{-28}^{+86}$   & 1334$-$127 & $121_{-142}^{+74}$ \\[0.5ex]
            0214$+$083 & $107_{-44}^{+63}$  & 1441$+$252 & $102_{-22}^{+46}$  \\[0.5ex]
            0250$-$225 & $-22_{-20}^{+106}$ & 1510$-$089 & $82_{-30}^{+12}$   \\[0.5ex]
            0301$-$243 & $154_{-85}^{+40}$  & 1520$+$319 & $158_{-18}^{+44}$  \\[0.5ex]
            0321$+$340 & $77_{-26}^{+59}$   & 1542$+$616 & $65_{-27}^{+39}$   \\[0.5ex]
            0430$+$052 & $199_{-110}^{+47}$ & 1553$+$113 & $75_{-112}^{+77}$  \\[0.5ex]
            0451$-$282 & $107_{-46}^{+77}$  & 1603$+$573 & $237_{-126}^{+22}$ \\[0.5ex]
            0506$+$056 & $234_{-79}^{+17}$  & 1633$+$382 & $75_{-23}^{+25}$   \\[0.5ex]
            0603$+$476 & $269_{-26}^{+22}$  & 1638$+$398 & $-5_{-17}^{+46}$   \\[0.5ex]
            0716$+$714 & $131_{-23}^{+13}$  & 1641$+$399 & $33_{-55}^{+43}$   \\[0.5ex]
            0722$+$145 & $271_{-12}^{+17}$  & 1700$+$685 & $139_{-72}^{+58}$  \\[0.5ex]
            0735$+$178 & $260_{-132}^{+27}$ & 1730$-$130 & $94_{-133}^{+88}$  \\[0.5ex]
            0738$+$548 & $161_{-41}^{+62}$  & 1749$+$096 & $142_{-84}^{+42}$  \\[0.5ex]
            0806$+$524 & $87_{-100}^{+101}$ & 1749$+$701 & $-33_{-12}^{+93}$  \\[0.5ex]
            0836$+$710 & $208_{-107}^{+17}$ & 1803$+$784 & $-14_{-19}^{+114}$ \\[0.5ex]
            0846$+$513 & $73_{-32}^{+119}$  & 1807$+$698 & $19_{-24}^{+94}$   \\[0.5ex]
            0906$+$015 & $122_{-22}^{+14}$  & 1828$+$487 & $122_{-110}^{+128}$\\[0.5ex]
            0917$+$449 & $181_{-54}^{+54}$  & 2155$+$312 & $20_{-22}^{+115}$  \\[0.5ex]
            0946$+$006 & $127_{-136}^{+21}$ & 2227$-$088 & $91_{-8}^{+18}$    \\[0.5ex]
            1101$+$384 & $47_{-87}^{+105}$  & 2233$-$148 & $21_{-18}^{+27}$   \\[0.5ex]
            1144$+$402 & $102_{-82}^{+104}$ & 2247$-$283 & $38_{-33}^{+83}$   \\[0.5ex]
            1156$+$295 & $5_{-18}^{+102}$   & 2251$+$158 & $66_{-100}^{+39}$  \\[0.5ex]
            1215$+$303 & $250_{-104}^{+12}$ & 2258$-$022 & $114_{-38}^{+38}$  \\[0.5ex]
            1222$+$216 & $248_{-27}^{+10}$  & 2319$+$317 & $-48_{-22}^{+26}$  \\[0.5ex]
            1226$+$023 & $270_{-17}^{+12}$  & 2342$-$161 & $93_{-22}^{+39}$   \\[0.5ex]
            1227$+$255 & $105_{-128}^{+16}$ &            &                    \\[0.5ex]
            \hline
    \end{tabular}
    \end{table}
    
    \begin{table*}
    	\centering
    	\caption{Derived distances between the central engine and the region of the $\gamma$-ray emission production $r_{\gamma,\text{proj}}$ (projected scale), $r_{\gamma}$ (de-projected scale) together with 1$\sigma$ uncertainties. 
    	We have used in the calculations core shift measurements $\Omega$ from \citet{2012A&A...545A.113P} and mark by asterisk additional three targets provided by Plavin et al.~(in prep.). Core shift measures obtained from the median value 125~$\mu$as \citep[adopted from][]{2012A&A...545A.113P} are marked by $\diamond$.}
    	\label{t:rgamma}
    	\begin{tabular}{c|c}
    	     \begin{tabular}{crrr}
            \hline
    		Source & $\Omega$ & $r_{\gamma,\text{proj}}$ & $r_{\gamma}$ \\
    		name & (pc GHz) & (pc) & (pc) \\
    		\hline
    		0110$+$318 & $15.53^\diamond$ & $0.93_{-0.81}^{+0.48}$ & $24.92_{-21.60}^{+12.97}$ \\ [0.5ex]
            0130$-$171 & $11.98^*$ & $0.21_{-1.22}^{+0.56}$ & $3.25_{-19.04}^{+8.78}$ \\ [0.5ex]
            0208$+$106 & $7.58^\diamond$ & $0.20_{-0.45}^{+0.30}$ & $1.32_{-3.05}^{+2.13}$ \\ [0.5ex]
            0214$+$083 & $3.66^\diamond$ & $-0.63_{-0.53}^{+0.38}$ & $-4.12_{-3.44}^{+2.41}$ \\ [0.5ex]
            0301$-$243 & $9.42^\diamond$ & $0.39_{-0.26}^{+0.29}$ & $0.61_{-0.55}^{+0.58}$ \\ [0.5ex]
            0321$+$340 & $1.14^*$ & $-0.47_{-0.43}^{+0.20}$ & $-2.76_{-2.48}^{+1.14}$ \\ [0.5ex]
            0430$+$052 & $0.85$ & $-0.96_{-0.24}^{+0.56}$ & $-5.46_{-1.33}^{+3.18}$ \\ [0.5ex]
            0451$-$282 & $18.93^\diamond$ & $0.86_{-0.86}^{+0.83}$ & $46.86_{-47.56}^{+45.86}$ \\ [0.5ex]
            0506$+$056 & $11.11^\diamond$ & $0.57_{-0.31}^{+0.31}$ & $0.97_{-0.86}^{+0.87}$ \\ [0.5ex]
            0716$+$714 & $10.16$ & $-1.15_{-0.23}^{+0.35}$ & $-60.97_{-12.09}^{+18.51}$ \\ [0.5ex]
            0806$+$524 & $5.56^\diamond$ & $0.22_{-0.23}^{+0.23}$ & $0.31_{-0.39}^{+0.39}$ \\ [0.5ex]
            0836$+$710 & $25.72$ & $0.55_{-0.53}^{+0.79}$ & $21.11_{-20.47}^{+30.76}$ \\ [0.5ex]
            0846$+$513 & $15.29^\diamond$ & $0.77_{-0.58}^{+0.42}$ & $11.13_{-8.56}^{+6.27}$ \\ [0.5ex]
            0906$+$015 & $29.45$ & $0.88_{-0.52}^{+0.54}$ & $24.69_{-14.48}^{+15.10}$ \\ [0.5ex]
            0917$+$449 & $19.47^\diamond$ & $0.68_{-0.56}^{+0.56}$ & $171.01_{-140.10}^{+140.17}$ \\ [0.5ex]
            0946$+$006 & $15.29^\diamond$ & $0.32_{-0.43}^{+0.85}$ & $36.59_{-48.70}^{+96.10}$ \\ [0.5ex]
            1101$+$384 & $2.81$ & $0.18_{-0.04}^{+0.04}$ & $4.42_{-1.28}^{+1.26}$ \\ [0.5ex]
            1156$+$295 & $20.11$ & $1.28_{-1.30}^{+0.50}$ & $28.17_{-28.59}^{+11.04}$ \\ [0.5ex]
            1215$+$303 & $5.28^\diamond$ & $0.23_{-0.14}^{+0.15}$ & $3.19_{-2.22}^{+2.33}$ \\ [0.5ex]
            1222$+$216 & $17.03$ & $-2.04_{-0.41}^{+0.53}$ & $-35.35_{-7.03}^{+9.02}$ \\ [0.5ex]
            1226$+$023 & $6.27^\diamond$ & $-2.50_{-0.22}^{+0.25}$ & $-58.79_{-5.01}^{+5.88}$ \\ [0.5ex]
            1236$+$049 & $19.84^\diamond$ & $1.34_{-0.53}^{+0.53}$ & $221.42_{-88.44}^{+87.87}$ \\ [0.5ex]
            1308$+$326 & $13.61$ & $-0.20_{-0.53}^{+0.51}$ & $-4.96_{-12.96}^{+12.53}$ \\ [0.5ex]
    		\hline
    	\end{tabular}
    	&
    	\begin{tabular}{crrr}
            \hline
    		Source & $\Omega$ & $r_{\gamma,\text{proj}}$ & $r_{\gamma}$ \\
    		name & (pc GHz) & (pc) & (pc) \\
    		\hline
    		1329$-$049 & $19.51^\diamond$ & $0.70_{-0.69}^{+0.77}$ & $1857.22_{-1829.02}^{+2050.57}$ \\ [0.5ex]
            1334$-$127 & $31.08$ & $0.63_{-0.97}^{+1.75}$ & $42.48_{-65.40}^{+117.76}$ \\ [0.5ex]
            1441$+$252 & $18.36^\diamond$ & $0.84_{-0.52}^{+0.50}$ & $30.86_{-19.11}^{+18.28}$ \\ [0.5ex]
            1510$-$089 & $13.50$ & $-0.53_{-0.38}^{+0.61}$ & $-12.00_{-8.65}^{+13.71}$ \\ [0.5ex]
            1520$+$319 & $19.82^\diamond$ & $1.23_{-0.53}^{+0.53}$ & $6.00_{-3.34}^{+3.34}$ \\ [0.5ex]
            1542$+$616 & $14.22^\diamond$ & $0.88_{-0.38}^{+0.38}$ & $8.62_{-4.23}^{+4.22}$ \\ [0.5ex]
            1603$+$573 & $16.78^\diamond$ & $-1.43_{-0.60}^{+1.46}$ & $-18.35_{-7.57}^{+18.68}$ \\ [0.5ex]
            1633$+$382 & $21.21$ & $0.73_{-0.58}^{+0.57}$ & $71.80_{-57.06}^{+56.36}$ \\ [0.5ex]
            1638$+$398 & $19.87^\diamond$ & $1.35_{-0.58}^{+0.54}$ & $15.18_{-6.65}^{+6.15}$ \\ [0.5ex]
            1641$+$399 & $23.85$ & $1.25_{-0.60}^{+0.70}$ & $85.51_{-41.17}^{+47.85}$ \\ [0.5ex]
            1700$+$685 & $10.29^\diamond$ & $-0.35_{-0.52}^{+0.60}$ & $-3.09_{-4.52}^{+5.29}$ \\ [0.5ex]
            1730$-$130 & $27.31$ & $0.67_{-1.19}^{+1.70}$ & $37.11_{-66.45}^{+94.84}$ \\ [0.5ex]
            1749$+$096 & $6.92$ & $-0.16_{-0.35}^{+0.47}$ & $-121.55_{-265.85}^{+359.32}$ \\ [0.5ex]
            1749$+$701 & $27.03$ & $1.86_{-0.49}^{+0.46}$ & $23.85_{-6.46}^{+6.11}$ \\ [0.5ex]
            1803$+$784 & $6.58$ & $0.50_{-0.68}^{+0.45}$ & $17.77_{-24.01}^{+15.84}$ \\ [0.5ex]
            1807$+$698 & $3.81$ & $0.25_{-0.06}^{+0.06}$ & $20.53_{-5.22}^{+5.20}$ \\ [0.5ex]
            1828$+$487 & $12.24$ & $-0.12_{-1.08}^{+0.96}$ & $-4.41_{-39.09}^{+34.78}$ \\ [0.5ex]
            2227$-$088 & $29.60$ & $1.65_{-0.54}^{+0.54}$ & $36.43_{-12.25}^{+12.19}$ \\ [0.5ex]
            2247$-$283 & $14.49^\diamond$ & $0.56_{-0.96}^{+0.52}$ & $7.15_{-12.28}^{+6.71}$ \\ [0.5ex]
            2251$+$158 & $22.00$ & $0.96_{-0.56}^{+0.91}$ & $66.06_{-38.70}^{+62.26}$ \\ [0.5ex]
            2258$-$022 & $17.27^\diamond$ & $0.32_{-0.54}^{+0.54}$ & $16.79_{-28.25}^{+28.39}$ \\ [0.5ex]
            2319$+$317 & $19.82^\diamond$ & $1.51_{-0.54}^{+0.54}$ & $12.27_{-4.50}^{+4.47}$ \\ [0.5ex]
            2342$-$161 & $15.74^\diamond$ & $0.69_{-0.45}^{+0.43}$ & $51.31_{-33.43}^{+32.13}$ \\ [0.5ex]
    		\hline
    	\end{tabular}
    	\end{tabular}
    	
    \end{table*}

    \subsection{Gamma-ray emission region: distance from the central engine}

    Having the time delays, we can estimate the distance $r_\gamma$ between the central engine and the region which dominates in the $\gamma$-ray emission production for each source independently. It is convenient to operate with projected distances $r_{\gamma,\text{proj}} = r_\gamma \sin{\theta}$. Then using \autoref{rg_dist} and \autoref{rc_dist}, we have
    \begin{equation}
        r_{\gamma,\text{proj}} = \frac{\Omega}{\nu} - \frac{\beta_{\text{app}} \Delta t_{\text{obs}}}{1 + z} \,.
        \label{rgamma_eq}
    \end{equation}
    The core shifts for 20 sources are measured by \citet{2012A&A...545A.113P} at the frequency pair of 8 and 15~GHz and have typical errors of 50~$\mu$as. We used core shift values for additional two AGNs from Plavin et al.\ (in prep.), they are measured between 2 and 8~GHz. For 24 more sources with calculated time delays but unknown core shifts, the value 125~$\mu$as \citep[the median from][]{2012A&A...545A.113P} is considered, which is justified by the fact that a core shift is often highly-variable \citep{2006ApJ...642L.115H,2019MNRAS.485.1822P}.
    
    The resulting distribution of $r_{\gamma,\text{proj}}$ (\autoref{rgamma_hist}, \autoref{t:rgamma}) comprises 46~values and has a median of 0.55~pc. To calculate the de-projected distances $r_\gamma = r_{\gamma,\text{proj}} / \sin{\theta}$, the viewing angles were taken from \citet{MOJAVE_XIX}. The values of $r_\gamma$ obtained in this way have a median of 11.7~pc which is well beyond the BLR. For 19 sources or 41\%, the de-projected distance $r_\gamma$ is greater than 1~pc with ${1\sigma}$ significance. This supports the scenario where the high-energy emission is produced at the distance of a few parsecs from the central engine. For 17 sources (37\%), $r_{\gamma}$ is indistinguishable from zero, and the $\gamma$-ray photons might originate at sub-parsec distances, or there are multiple mechanisms responsible for the $\gamma$-ray emission production. We also found considerable difference between the distances typical for the two source classes: 24.7~pc for the quasars (29 sources) and 1.3~pc for the BL Lacs (13 sources).
    
    Note that there are a few sources with negative estimated distances; it has no meaningful physical interpretation. Leaving aside the above-mentioned core shift variability, it is likely that the negative distances arise from the time lags between physically unrelated radio/$\gamma$-ray flares. Alternatively, it can be explained by the fact that \autoref{rgamma_eq} does not take into account that (i) there is a transition from parabolic to conical shape in the jet \citep{2020MNRAS.495.3576K} -- backward extrapolation of the conical jet makes it shorter, (ii) apparent plasma speed $\beta_{\text{app}}$ within the 15~GHz core is likely lower than the one measured beyond it, as the jets still undergo a residual acceleration on these scales \citep{2015ApJ...798..134H}. Thus, in some cases \autoref{rgamma_eq} estimates a lower limit on $r_{\gamma,\text{proj}}$, though the true value is not expected to be significantly larger.
    
    \subsection{Gamma-ray emission region: size estimation}
    
    \begin{figure}
        \centering
        \includegraphics[width=\linewidth]{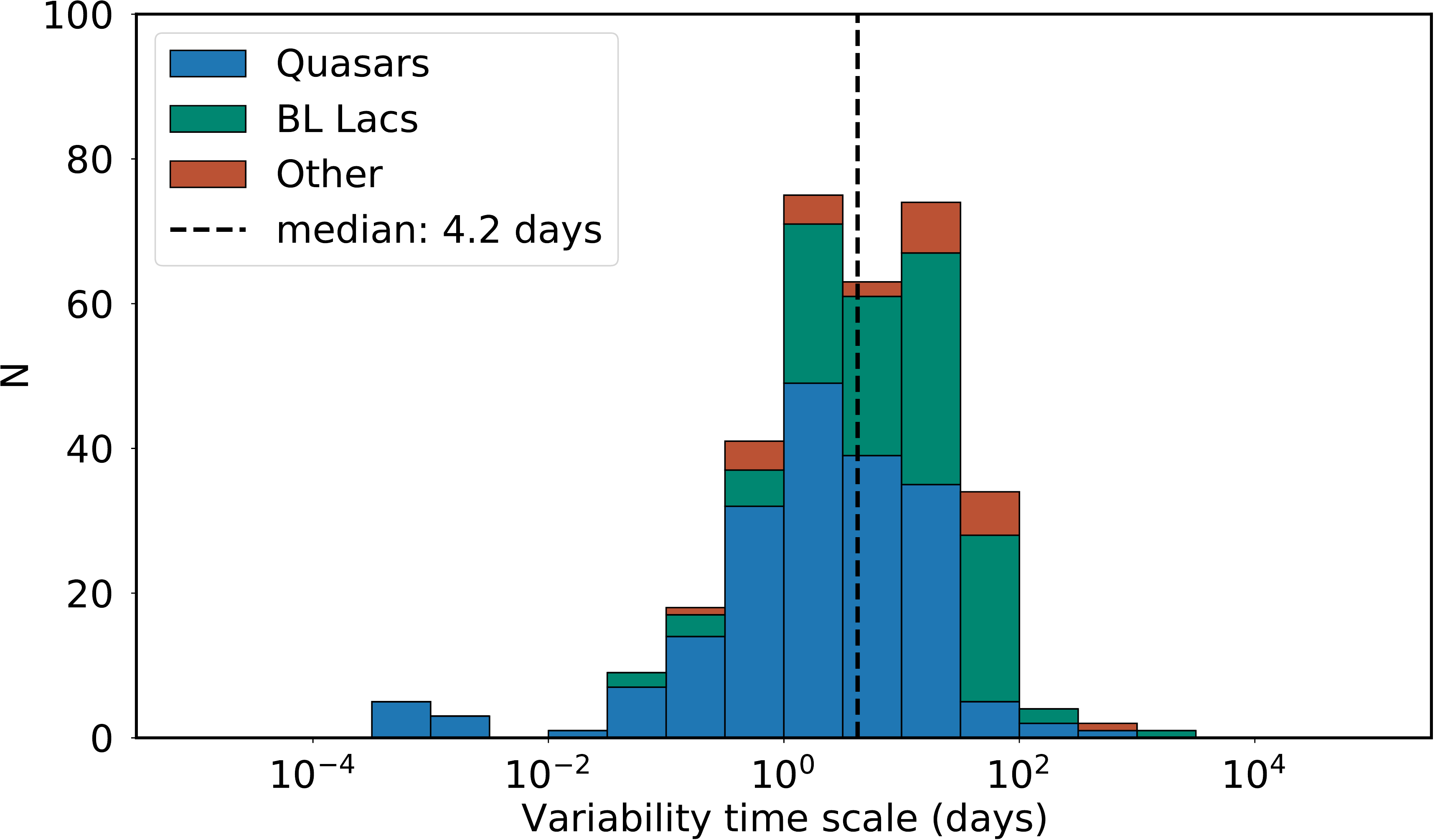}
        \caption{Distribution of the variability time scales obtained from the adaptive binned light curves in the observer's frame (330 sources). The blue, green and orange bars refer to the quasars, BL Lacs and other source classes, respectively.}
        \label{flares_hist}
    \end{figure}
    
    \begin{figure}
        \centering
        \includegraphics[width=\linewidth]{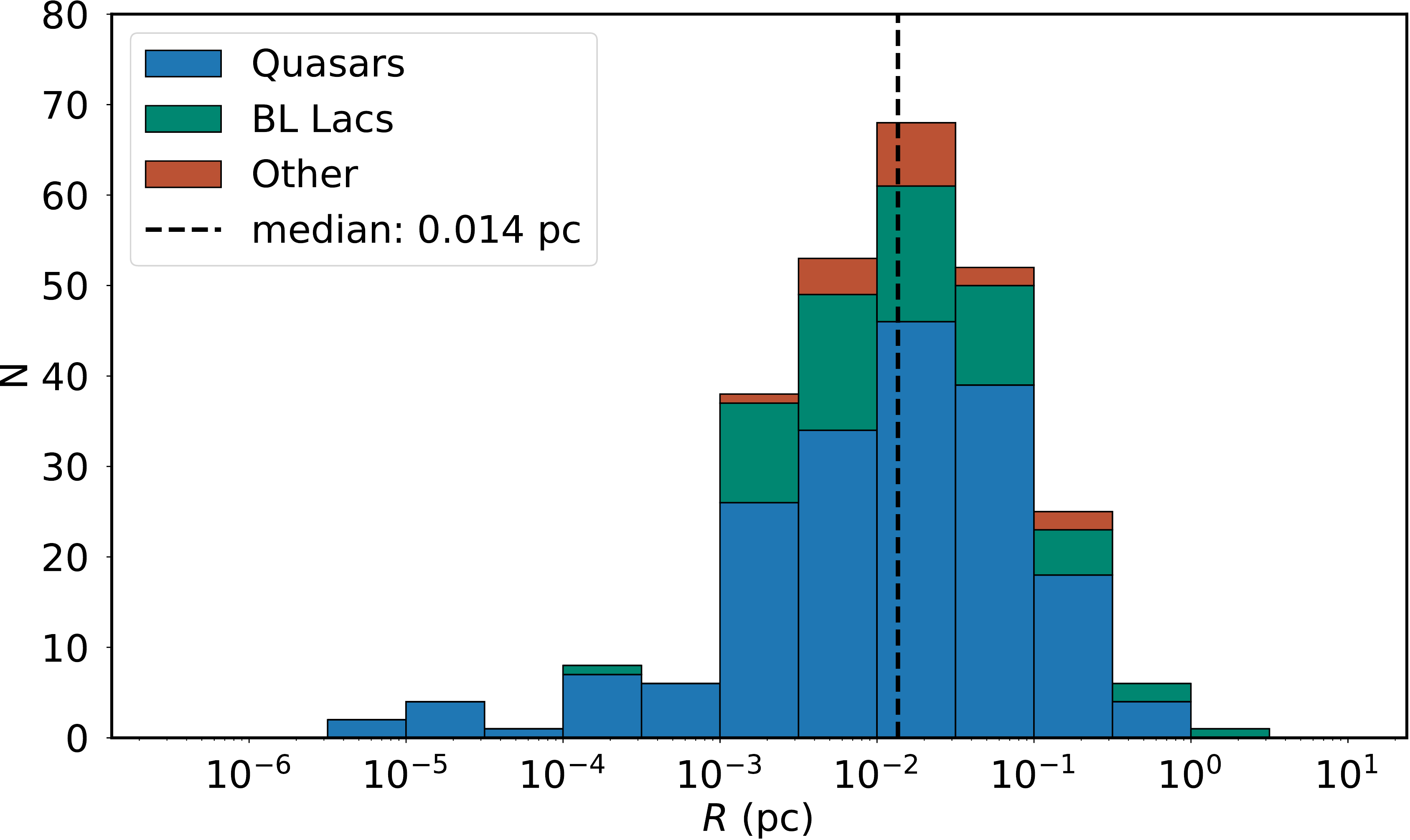}
        \caption{Distribution of the $\gamma$-ray emission region transverse sizes in the plasma frame estimated from the causality argument (264 sources). The blue, green and orange bars refer to the quasars, BL Lacs and other source classes, respectively.}
        \label{gamma_size_hist}
    \end{figure}
    
    An upper limit on the size of the $\gamma$-ray emission region $R$ in the jet co-moving frame can be estimated from the causality argument:
    \begin{equation}
        R \lesssim \frac{c t_\text{var} \delta}{1 + z},
    \end{equation}
    where $t_\text{var}$ is the observed $\gamma$-ray variability time scale, and $\delta$ is the Doppler factor. To obtain a typical $t_\text{var}$ value, we follow the method suggested by \citet{2019ApJ...877...39M}. In this section, we studied the adaptive binned light curves with no restrictions on the included data points. In the first step, a $\gamma$-ray flare peak is defined as a flux block which is higher than the previous and subsequent blocks. Here, the block is understood as a single data point. Then, all the successively lower blocks on either side of the peak are included in the flare regardless of the flux change in comparison to the flux of the peak. If the minimum duration of a flare rise phase is taken as a variability time scale, the procedure yields a median of 4.2~days. The variability time scale of 75\% of the sources falls in the range of 1 to 100 days, but there was also found a considerable number of extremely short flares of a duration of a few hours (\autoref{flares_hist}). The difference between the source classes is also present in this histogram (the median is 2.2~days for the quasars and 11.0~days for the BL Lacs).
    
    Finally, we used Doppler factors from \citet{MOJAVE_XIX} and obtained the $\gamma$-ray emission region size $R \lesssim 0.014$~pc (see \autoref{gamma_size_hist}). It is typically a factor of 20-50 times smaller than the jet width at the corresponding distances \citep[e.g.][Figure~1]{2020MNRAS.495.3576K}, and about two orders of magnitude more compact than the transverse size of the 15~GHz core \citep{PK15_scattering}. If we consider a longitudinal dimension of the synchrotron self-absorbed core, which has a cigar-like shape, the ratio will further decrease, explaining considerably longer radio flares, compared to those in $\gamma$-rays. The difference between the transverse sizes for the quasars and BL Lacs was found negligible (0.14 and 0.12~pc, respectively).
    
    We also note that the initial stages of a radio flare at mm wavelengths might coincide with the $\gamma$-ray high states \citep{2011A&A...532A.146L}. Though it is not easy to determine the time when the radio flare begins, we do see such examples at the cm radio domain as well (see \autoref{light_curve}), i.e. when the 15~GHz radio flux has already started to rise before the $\gamma$-ray flaring starts. It further argues for $\gamma$-rays originating relatively far from the central engine.

\section{Summary}
\label{sec:conclutions}

We have studied the correlation between the 15~GHz VLBA flux densities and 100~MeV -- 300~GeV $\gamma$-ray photon flux of 331 $\gamma$-bright \textit{Fermi} blazars using the ZDCF correlation method. We come to the following conclusions.
\begin{enumerate}
    \item There is a significant correlation between the $\gamma$-ray photon flux and the 15~GHz VLBA core flux density. A typical delay is about 3--5~months in the observer's frame and of 2--3~months in the source frame; the $\gamma$-ray emission precedes radio. This indicates that the $\gamma$-ray emission is produced between the vicinity of the black hole and the 15~GHz core.
    \item The correlation is more pronounced when (i) the adaptive $\gamma$-ray light curves are used for the stacking analysis or (ii) the brightest in $\gamma$-rays half of the sample is considered. This suggests that short ($\lesssim 1$~week) and strong $\gamma$-ray flares are blurred in the weekly binned light curves.
    \item A significant correlation with the radio emission of downstream parsec-scale jet regions is found at greater time lags of 5--9~months than with the core component. This is a result of moving plasma or developing instabilities being emerged from the radio core and propagated through the innermost, often quasi-stationary, VLBI jet features.
    \item
    For 46 sources we derived the de-projected distance between the central engine and the region of the $\gamma$-ray emission. For 19 of them, it is found to be greater than 1~pc, with significance exceeding $1\sigma$.
    The median distance over the 46 sources is estimated about 11.7~pc. We conclude that the seed photons responsible for the $\gamma$-ray emission are likely to originate beyond the broad line region, on scales where jets undergo active collimation and acceleration.
    \item The quasars have on average more significant ZDCF peak, more distant $\gamma$-ray emission region and shorter variability time scale than the BL Lacs.
\end{enumerate}

\section*{Acknowledgements}

We thank Daniel Homan, Eduardo Ros for productive discussions and comments, and Elena Bazanova for language editing. We thank the anonymous referee for useful comments that helped to improve the manuscript. This study has been supported by the Russian Science Foundation grant 20-72-10078. ABP and TS were supported in part by the Academy of Finland projects 296010 and 318431, 315721, respectively. TH was supported by the Academy of Finland projects 317383, 320085, and 322535. This research has made use of data from the MOJAVE database that is maintained by the MOJAVE team \citep{2018ApJS..234...12L}. The MOJAVE project was supported by NASA-Fermi GI grants NNX08AV67G, NNX12A087G and NNX15AU76G. This research has made use of NASA's Astrophysics Data System Bibliographic Services.

\section*{Data Availability}

Raw \textit{Fermi} LAT data used to construct the light curves are available from the \textit{Fermi} Data Server\footnote{\url{http://heasarc.gsfc.nasa.gov/FTP/fermi/data/lat/weekly/diffuse}}. The $\gamma$-ray light curves in machine readable form will be shared on reasonable request to the corresponding author. 
Fully calibrated 15~GHz VLBA images and visibility data obtained mainly within the MOJAVE program and in some part from the NRAO archive experiments reduced by us are available online\footnote{\url{http://www.physics.purdue.edu/astro/MOJAVE/allsources.html}}.
VLBI source structure modelfits containing the core and jet feature flux densities are available from \citet{MOJAVE_XVIII}.
Our Python 3 code on ZDCF is publicly available and can be found on GitHub\footnote{\url{https://github.com/hey-moon/ZDCF}}.

%



\bibliographystyle{mnras}
\bibliography{main}







\bsp	
\label{lastpage}
\end{document}